\documentclass[12pt]{article}
\usepackage{latexsym}
\usepackage{epsfig,amssymb,euscript,slashed}
\usepackage{amsmath}

\usepackage{array,calc,epsfig}
\usepackage[linktocpage]{hyperref}
\usepackage{pstricks}
\usepackage{bbm}
\usepackage{fancybox}
\usepackage{hyperref}
\usepackage{cite} 

\def\be{\begin{equation}}
\def\ee{\end{equation}}
\def\bseq{\begin{subequations}}
\def\eseq{\end{subequations}}

\def\bea{\begin{eqnarray}}
\def\eea{\end{eqnarray}}
\newcommand\bbone{\ensuremath{\mathbbm{1}}}
\newcommand{\ul}{\underline}
\def\bseq{\begin{subequations}}
\def\eseq{\end{subequations}}

\numberwithin{equation}{section} 
\usepackage[]{graphicx}

\def\d {{\rm d}}

\def\calb         {{\cal B}}
\def\calc         {{\cal C}}

\def\cali         {{\cal I}}

\def\call         {{\cal L}}
\def\calm         {{\cal M}}
\def\caln         {{\cal N}}
\def\calo         {{\cal O}}

\def\calq         {{\cal Q}}
\def\calr         {{\cal R}}

\def\calt         {{\cal T}}

\def\calv         {{\cal V}}

\def\del          {\partial}
\def\delbar       {\bar\partial}
\def\ii           {{\rm i}}

\def\Re           {{\rm Re\hskip0.1em}}
\def\Im           {{\rm Im\hskip0.1em}}

\def\sqr#1#2{{\vcenter{\vbox{\hrule height.#2pt
 \hbox{\vrule width.#2pt height#1pt \kern#1pt \vrule width.#2pt}\hrule
 height.#2pt}}}}

\newcommand{\ft}[2]{{\textstyle{\frac{#1}{#2}}}}

\def\d{\text{d}}

\def\slashchar#1{\setbox0=\hbox{$#1$}           
\dimen0=\wd0                                 
\setbox1=\hbox{/} \dimen1=\wd1               
\ifdim\dimen0>\dimen1                        
\rlap{\hbox to \dimen0{\hfil/\hfil}}      
#1                                        
\else                                        
\rlap{\hbox to \dimen1{\hfil$#1$\hfil}}   
/                                         
\fi}

\newcommand{\Pint}{\mathbb P}
\newcommand{\Rint}{\mathbb R}
\newcommand{\Zint}{\mathbb{Z}}
\newcommand{\Cint}{\mathbb{C}}

\newcommand{\ba}{\begin{array}}
\newcommand{\ea}{\end{array}}
\def\nn{\nonumber}

\begin{document}
\font\cmss=cmss10 \font\cmsss=cmss10 at 7pt

\vskip -0.5cm
\rightline{\small{\tt ROM2F/2012/06}}

\vskip .7 cm

\hfill
\vspace{18pt}
\begin{center}
{\Large \textbf{Branes, U-folds and hyperelliptic fibrations}}
\end{center}

%
\vspace{6pt}
\begin{center}
{\textsl{ Luca Martucci\footnote{\scriptsize \tt luca.martucci@roma2.infn.it}, Jose Francisco Morales\footnote{\scriptsize \tt francisco.morales@roma2.infn.it} \& Daniel Ricci Pacifici \footnote{\scriptsize \tt daniel.ricci.pacifici@roma2.infn.it}}}

\vspace{1cm}
\textit{\small I.N.F.N. Sezione di Roma ``TorVergata'' \&\\  Dipartimento di Fisica, Universit\`a di Roma ``TorVergata", \\
Via della Ricerca ScientiÞca, 00133 Roma, Italy }\\  \vspace{6pt}
\end{center}

\begin{center}
\textbf{Abstract}
\end{center}

\vspace{4pt} {\small

\noindent We construct a class of supersymmetric vacua of type IIB  string theory describing
systems of  three- and seven-branes non-perturbatively completed by brane instantons.   
 The vacua are specified by a set of holomorphic functions defined over a  complex plane 
 up to non-trivial  U-duality monodromies around the brane locations.  
 In the simplest setting, the solutions can be seen as a generalization of F-theory elliptic fibrations, where the torus fiber
 is replaced by a genus two Riemann surface with periods encoding the information on 
  the axio-dilaton, the warp factor and the NS-NS and R-R fluxes.  
  
\noindent }
%

%
%
%
%
%






\newpage

\tableofcontents

\section{Introduction and Summary}

F-theory \cite{Vafa:1996xn} provides an elegant framework where fully non-pertubative solutions of type IIB supergravity 
are described in purely geometric terms. Solutions are characterized by a non-trivial profile of the axio-dilaton field $\tau$.
   In the simplest set up, one can think of the field $\tau$ as the complex structure of an auxiliary torus fibered over a complex plane with punctures  at the points where the torus fiber degenerates.  
       Moving around a puncture, $\tau$ undergoes a non-trivial monodromy  in the U-duality group ${\rm SL}(2,\mathbb{Z})$
       indicating the presence of a 7-brane charge. The resulting vacua are non-perturbative in nature 
 but in appropriate limits they describe systems of D7-branes and O7-planes non-perturbatively completed by D-instantons 
 \cite{sen1,sen2}. In particular, an O7-plane is described in this framework as a composite object, a pair of ($p,q$) 7-branes 
 colliding at weak coupling.    In this paper we study an extension of this picture which includes 3-branes as well. 
   
The general philosophy is the same as the one adopted in F-theory, with the difference that now we start from type IIB theory on K3 and consider solutions with non-trivial profiles on a complex plane for a set of scalar fields in the six-dimensional effective theory.
The effective six-dimensional  supergravity   includes 105 scalars, which are rotated by an ${\rm O}(5,21;\mathbb{Z})$ U-duality group. 
  By allowing a subset of these scalars to vary over a complex plane with non-trivial U-duality monodromies, we will construct supersymmetric solutions of the six-dimensional supergravity, {\it U-folds},  that incorporate within the same framework 3- and 7-branes \footnote{See for instance \cite{Kumar:1996zx,Liu:1997mb,Hellerman:2002ax,Flournoy:2004vn,Gray:2005ea,Vegh:2008jn,McOrist:2010jw} for previous work discussing similar extensions of F-theory.}.   
  The  solutions generically describe (from the ten-dimensional perspective) non-geometric string vacua which patch together mutually non-local systems of 3- and 7-branes.  In analogy with the F-theory case,  the U-folds we consider here 
    can be interpreted, in some appropriate limits, in terms of systems of D3,D7-branes and O3,O7-planes complemented by 
    D(-1) and  ED3 instantons. 
    
   We focus on solutions preserving $\caln=2$ four-dimensional supersymmetry. This allows us to make contact  with 
   field-theoretic results   based on the Seiberg-Witten analysis \cite{Seiberg:1994rs} and their M-theory engineering 
     \cite{Witten:1997sc}. In \cite{cremonesi}, the supergravity vacuum associated with a system of fractional D3-branes at a $\Cint^2/\mathbb{Z}_2$ singularity was obtained by reduction  of the M5 brane solution along a two-dimensional curve.   More recently, in \cite{lerdaetnoi}, this solution  was derived directly from string amplitudes computing the rate of emission  of twisted fields from fractional D3-branes and D-instanton sources at the singularity.  In particular, the profile for the twisted field on the plane orthogonal to the singularity was related to certain chiral correlators in the dual gauge theory (see also 
\cite{Billo:2010mg,Billo:2011uc,Fucito:2011kb} for similar results in the elliptic F-theory context). 
To understand, the gravity counterpart of these results was one of the initial motivations of this work. 
Here we develop a unifying framework for supergravity solutions describing general systems  (geometric or not) of 3- and 7-branes, in which brane instanton corrections  are codified in simple geometrical terms.  Although we confine ourselves to ${\cal N}=2$ supersymmetric vacua, we believe our techniques can be adapted to less supersymmetric and phenomenologically motivated settings, as for instance those of \cite{Kachru:2003aw,Balasubramanian:2005zx}.

  
  Let us now briefly discuss  our approach and the structure of the paper. In section \ref{sechol}, we review the construction of ten-dimensional holomorphic vacua of type IIB supergravity. We show that after reduction on K3, the conditions
  of ten-dimensional supersymmetry translate into the requirement  of holomorphicity for a set of six-dimensional fields
  $(\tau,\sigma,\beta^a)$.   $\tau$ is the axio-dilaton field, $\sigma$ characterizes the warp factor and the R-R four-fom  and 
  $\beta^a$ correspond to the reduction of the NS-NS/R-R two-form $C_2+\tau B$ along a set of $n$ vanishing exceptional cycles $\calc_a$ at  a singularity of K3.  
    These   fields  transform  under an  ${\rm O}(2,2+n;\mathbb{Z})$ subgroup of the complete  U-duality group. 
   The six-dimensional viewpoint is developed in section \ref{sec:IIBcomp}, where the ten-dimensional equations 
   (integrated over K3) are reproduced  directly from the six-dimensional effective theory, along the lines of \cite{cstring}. This provides a framework in which the global properties of the vacua, characterized by non-trivial monodromies in the U-duality group, can be addressed. 
  
In section \ref{sec:examples1}, \ref{sec:examples2}  we provide some explicit realizations of our general results. The case with no three-form fluxes ($n=0$) is described by a double elliptic fibration over a complex plane, a double copy of  the well understood F-theory  elliptic geometries.  Already for $n=1$ (associated with a $\Cint^2/{\mathbb{Z}}_2$ singularity) one obtains  a much richer situation. The U-duality group in this case is  ${\rm O}(2,3;\mathbb{Z})\simeq {\rm Sp}(4,\Zint)$ which is nothing but the modular group of a genus two Riemann surface. Moreover, one can see that $\tau,\sigma,\beta$ transform under this group as the three entries of the period matrix ${\bf \Omega}$ of a genus two surface. This suggests that the general solution in this case is described by the fibration of a genus two  surface over $\Cint$ such that ${\bf \Omega}(z)$ varies holomorphically on $z$. Locations of branes are associated with points in 
the $z$-plane where the fiber degenerates. Circling  these points, the matrix ${\bf \Omega}(z)$ undergoes non-trivial 
U-duality monodromies specifying the type of brane at the puncture. 
We discuss some simple examples and their brane interpretation.
A more systematic analysis, and extensions to cases with $n>1$, is left to future investigation.
Finally, this paper contains extensive appendices including technical details and background material.   

%

\section{Holomorphic solutions of type IIB supergravity}
\label{sechol}

We are interested in describing type IIB vacua preserving $\caln=2$ four-dimensional supersymmetry and characterized by the presence of D3 and D7-branes. We start with a ten-dimensional background of the form $\mathbb{R}^{1,3}\times \mathbb{C}\times X$, where $X$ is a (four-dimensional Ricci-flat) K3 space. We consider  D-branes with world-volumes sharing   the four-dimensional flat space $\mathbb{R}^{1,3}$ and sitting at certain  points in $\mathbb{C}$. Hence, D7's wrap the entire internal $X$ while D3's sit at  points in $X$.   
Furthermore D3-branes can be either regular or fractional in the case K3 is singular. 

These configurations preserve $\caln=2$ four-dimensional supersymmetry and locally admit a specific ten-dimensional 
  supergravity description. Indeed,  they can be seen as special sub-cases of the warped Calabi-Yau/F-theory backgrounds discussed in \cite{GKP},  which considers warped flux vacua on space-times of the form $\mathbb{R}^{1,3}\times Y$, where $Y$ is either a Calabi-Yau or a F-theory K\"ahler space. In our case we locally have $Y=\mathbb{C}\times X$. The general backgrounds of \cite{GKP} preserve $\caln=1$ four-dimensional supersymmery, which is enhanced to $\caln=2$ for this choice. These kinds of configurations have been previously considered in the literature, as for instance in \cite{GP}.

 \subsection{Supersymmetric vacua}

In this section we review the construction in  \cite{GP} of holomorphic vacua of type IIB supergravity on a singular K3
describing the local geometry generated by  a systems of  D3 and D7 branes.
In Appendix \ref{app:10d} we present a self-contained re-derivation of these solutions by using the generalized complex geometry formalism.  Here we just quote the results.
 The ten-dimensional metric in the Einstein frame is given by
 \be\label{10dmetric}
 \d s^2_{\rm E} =  e^{2 A}\d x^\mu\d x_\mu+ e^{-2A} \d s^2_Y
  \ee
 with
 \be\label{ymetric}
\d s^2_Y=  e^{-\phi} 
 |h(z)|^2 \d z \d\bar z +\d s^2_{X} 
\ee
Here $\d s^2_X$ is the Ricci flat K3 metric, $h(z)$ is a holomorphic function and the dilaton $\phi$ is constant along $X$.
In addition, two-form potentials are taken to be self-dual with respect to the K3 metric defined by hyperk\"ahler structure associated with the triplet of anti-self dual two-forms $(\Re\omega,\Im\omega,j)$, where $j$ is the K\"ahler form 
and $\omega$ the holomorphic (2,0) form  on $X$.
 By introducing a set $\chi_a\in H^2(X;\mathbb{Z}) $ of integer  self-dual two-forms on $X$,
with positive definite non-degenerate pairing
\be\label{deltab}
\Delta_{ab}=\int_X\chi_a\wedge\chi_b
\ee
we set
\be
C_2+\tau B=\beta^a\chi_a
\ee 
 We remark that self-duality of $\chi_a$
implies $j\wedge\chi_a=\omega\wedge\chi_a=\bar\omega\wedge\chi_a=0$ and therefore the two-cycles $\calc_a$
 Poincar\'e dual to the forms $\chi_a=[\calc_a ]$ should have vanishing volume\footnote{\label{foot:Pduality}We use the following definition of Poincar\'e duality: a two-form $[\calc_a]$  is Poincar\'e dual to a two-cycle $\calc_a$ if $\int_{\calc_a}\alpha=-\int_X\alpha\wedge [\calc_a]$, for any two-form $\alpha$. 
 We have $\int_X [\calc_a]\wedge [\calc_b]=-\calc_a\cdot \calc_b$, with $\calc_a\cdot \calc_b$ being the ordinary  intersection number of cycles.}.   
Indeed, as we will more precisely discuss in section \ref{sec:Dbranes}, non-trivial two-form fluxes  $\beta^a$ are allowed only for a singular K3 and signal for D5 branes wrapping the vanishing exceptional cycles $ \calc_a$ or 
 equivalently fractional D3-branes. The functions $\beta^a$ are taken to be constant on $X$ and varying over the $z$-plane.

Under these assumptions, the background supersymmetry conditions reduce drastically and can be written in the compact
form  (see Appendix \ref{app:10d} for details)
 \be
  \bar\partial {\cal T} +  \partial\bar {\cal T}=0     \label{susytau}
  \ee
  with $\bar\partial=\bar\partial_{\Cint}+\bar\partial_{X}=\d\bar u^\alpha \frac{\partial}{\partial \bar u^\alpha}$ (where $u^\alpha=(z,u^1,u^2)$, are local complex coordinates on $Y$) and $\cal T$ a polyform
\be
 {\cal T}=e^B\wedge  \, \left[C+\ii e^{-\phi}   \cos( e^{ {\phi\over 2}-2A} J ) \right] \label{tpoly}
  \ee
 packing the RR potentials $C=C_0+C_2+C_4$ and the NS-NS data. In particular
   \be
  J=j-\frac{\ii}{2} e^{-\phi} |h(z)|^2 \d z\wedge \d\bar z
  \ee
  is just the K\"ahler form associated with the metric (\ref{ymetric}). Writing $\calt$ in components
  \be
  \begin{aligned}
   {\cal T}_0 &= \tau\equiv C_0+\ii \, e^{-\phi}  \\
   {\cal T}_2  &= C_2 + \tau B =\beta^a\,\chi_a\\
    {\cal T}_4 &= C_4-  \ft{\ii}{2} e^{-4A}J\wedge J   
   +C_2\wedge B +  \ft12   \tau  B\wedge B  \label{t024}
    \end{aligned}
    \ee
one can immediately see that the first two conditions encoded in (\ref{susytau}) just require that we must take $\tau$ and $\beta^a$ to be holomorphic: $\tau=\tau(z)$ and $\beta^a=\beta^a(z)$.    

The last equation in (\ref{susytau}) requires more care. This can be seen from the integrability condition $\partial \bar\partial {\rm Re}{\cal T}_4=\partial \bar\partial {\rm Im}{\cal T}_4=0$. For instance the second equation implies  
\be
\del\delbar (e^{-4A}) \wedge J\wedge J=\frac{G_3\wedge \bar G_3}{2\,\Im\tau}    \label{lapwarp}
\ee
with
\be
G_3=\d C_2+\tau \,\d B= \big(\del\beta^a-\frac{\Im\beta^a}{\Im\tau}\,\del\tau\big)\wedge\chi_a
\ee
Equation (\ref{lapwarp}) determines  the warp factor $e^{-4A}$  in terms of the three form flux $G_3$. Notice that the right
hand side of this equation is localized at the singularity, which implies that $e^{-4A}$ should  depend on  $X$ in   order to match this behavior. A similar equation and conclusion can be drawn for  $C_4$.
   Still, one can define a holomorphic field on $\mathbb{C}$ out of ${\cal T}_4$. Indeed, by integrating (\ref{susytau}) over $X$ one gets
   \be
   \Re(\bar\partial_{\Cint} \sigma)=0      \qquad   {\rm with}  \qquad \sigma\equiv\int_X {\cal T}_4
   \ee
  which implies that $\sigma(z)$ is holomorphic on $\Cint$.   
  
   We can summarize these results by saying that the effective six-dimensional fields which are obtained by integrating $\calt$ along the internal 0-, 4- and 2-cycles
  \be
  \begin{aligned}
   \tau &=  {\cal T}_0\equiv C_0+\ii e^{-\phi } \\
 \sigma &=    \int_{X} {\cal T}_4 =   \int_{X} \left(C_4-\ft{\ii}{2} e^{-4A}  J\wedge J+B\wedge C_2+\ft12 \tau B \wedge B \right) \\
   \beta_a &= \int_{\calc_a} {\cal T}_2= \int_{\calc_a}  (C_2+\tau B) \equiv -\Delta_{ab}\beta^b   \label{sols0}
 \end{aligned}
 \ee
 depend holomorphically on $z$, i.e.
\be\label{holcond0}
\delbar\tau=0\, ,\quad \delbar\sigma=0\, ,\quad \delbar\beta^a=0
\ee
This result will be confirmed by  purely effective six-dimensional arguments in section \ref{sec:IIBcomp}.

 \subsection{D-branes and monodromies}
\label{sec:Dbranes}

 In presence of D-branes, the equations  (\ref{holcond0}) are modified by delta-like functions centered at the brane positions.
 Let us consider D$p$-branes, $p=3,5,7$, filling $\mathbb{R}^{1,3}$,  sitting at a point $0\in\Cint$ and 
 wrapping a $(p-3)$-cycle $\Sigma_{p-3}$ in $X$.  
 The elementary brane couples to the R-R fields via the CS term $\int_{\mathbb{R}^{1,3}\times \Sigma_{p-3}}C_{p+1}$. This coupling generates a source term of the internal R-R field strength 
\be
\d (e^B\wedge F)=\delta_{\mathbb{C}}^{2}(0)\wedge  \alpha_{7-p}   
\ee
where $F=F_1+F_3+F_5$, with $F_k$ the R-R field strengths and  $\alpha_{7-p}$ stands for the delta-like form  which
is Poincare dual in $X$ to  $\Sigma_{p-3}$.
  Integrating this equation over  $D_2\times [\alpha_{7-p}]$, where $D_2$ is a disk surrounding $0\in\Cint$, we get
   \be
 1=\oint_\gamma\int_{[\alpha_{7-p}]} e^B\wedge F =\int_{[\alpha_{7-p}]}   (e^B\wedge C) \Big|^{z\, e^{2\pi i}}_{z}  \label{uno}
 \ee
 where $\gamma=\partial D_2$ is a curve surrounding   $0\in\Cint$. In deriving the last equality 
 we have used the R-R Bianchi identity $\d_H F=0$ to write locally $e^B\wedge F=\d(e^B\wedge C)$.
 Specifying to $p=3,5,7$, we notice that the quantities in the right  hand side of (\ref{uno}) are nothing but the real parts of 
 $\sigma,\beta^a$ and $\tau$ respectively as defined 
 in (\ref{sols0}).  Then, equation (\ref{sols0}) implies  that the presence of D$p$-branes induces the monodromies \footnote{\label{footind}
  Curvature corrections for D7-brane wrapping a K3 surface induces a -1 unit of D3-brane charge  \cite{Bershadsky:1995sp}.
  What we refer here as a D7-brane is better thought as a D3D7 bound state wrapping K3 with zero net D3-brane charge  
  \cite{hori97}.
  }  
 \bea
 \text{D3}:&&\qquad \sigma\rightarrow \sigma+1\nn\\
 \text{D5$_a$}:&&\qquad \beta^b\rightarrow \beta^b+\delta^b_a     \label{monD3} \\
 \text{D7}:&&\qquad \tau\rightarrow \tau+1\nn
 \eea
    A similar analysis can be done for the holonomies associated with O-planes.
    Still, from the experience of F-theory it is known that O-planes are not elementary objects and at the non-perturbative level they  are resolved into  more elementary ones. Therefore the solutions will be characterized entirely in terms of the D-branes discussed above and their U-duals, which provide the elementary constituents of our background.


Finally, let us observe that $\tau,\sigma$ and $\beta^a$ can be also interpreted as the tree level 
complexified gauge couplings $\tau_{\rm YM}=\frac{\theta_{\rm YM}}{2\pi}+\frac{4\pi\ii}{g^2_{\rm YM}}$ appearing in the   
  the four-dimensional effective theories supported by the different branes probing these backgrounds. 
Indeed, by dimensionally reducing the DBI+CS action, one gets the identifications
$\tau^{\rm D3}_{\rm YM}= \tau$, $\tau^{\text{D7-D3}}_{\rm YM}=\sigma$, $\tau^{{\rm D5}_a}_{\rm YM}=\beta_a\equiv -\Delta_{ab}\beta^b$. The non-trivial profiles in the $\Cint$-plane for $(\sigma,\beta^a,\tau)$ in the gravity solutions we will construct describe then the running of these gauge couplings in the dual gauge theories.

  \subsection{U-dualities  }

So far, we have only considered elementary D-branes but one could consider other branes which are related to the above ones by duality transformations.  We are interested
in duality transformations whose action closes on fields $\tau,\sigma,\beta^a$ which characterize our vacua. 

Already at the level of ten-dimensional supergravity, one has the perturbative dualities which correspond to integral shifts of the R-R and NS-NS gauge potentials, as well as the non-perturbative  type IIB S-duality. In addition, in the compactified theory one has  an additional 
duality, the so called Fourier-Mukai transform that we denote by  $R$. 
This action does not has a counterpart in ten-dimensional supergravity. 
In our context, it can be seen as a sort of `T-duality' (along all four directions of the K3 space $X$) which exchanges regular D3 and D7-branes, leaving fractional D3-branes untouched. 
More precisely, the $R$-duality exchanges $\tau\leftrightarrow \sigma$ \footnote{We recall that the shift monodromy in $\tau$ is associated to a D3-D7 bound state wrapping K3 with zero net D3 brane charge, see footnote \ref{footind}. The fact that the bound state is dual to a D3-brane on K3  is supported by the analysis in  \cite{hori97}, where the two moduli spaces were matched.}. We can now combine the Fourier-Mukai tranform $R$ with S-duality and shift dualities  obtaining the following minimal set of duality tranformations acting on the fields $\tau,\sigma,\beta^a$ as follows:
\be\label{duality}
\begin{tabular}{|c|c|}
\hline
generator & non-trivial action\\
 \hline\hline
 $S$ & $\tau\rightarrow -\frac{1}{\tau} \qquad~  \sigma\rightarrow \sigma-\frac{1}{2\tau}\Delta_{ab}\,\beta^a\beta^b     \qquad~ \beta^a\rightarrow \frac{1}{\tau}\beta^a$ \\
  $T$ & $\tau\rightarrow \tau+1$\\
 $W_{a}$ & $\quad \beta^b\rightarrow \beta^b+\delta^b_a $\\
 $ R $ & $\tau\leftrightarrow \sigma$\\
 \hline
\end{tabular}
\ee
  The elements $S$ and $T$ generate the SL(2,$\mathbb{Z})_\tau$  S-duality group of type IIB theory, $R$ is the Fourier-Mukai transform and $W_a$  corresponds to the axionic shift $C_2\rightarrow C_2+\chi_a$
  \footnote{\label{Bshift}Alternatively, instead of $W_a$,  one can use as generators the conjugates $\tilde W_a=S^3\, W_{a} \,S$, acting as  shifts of the $B$-field: $\tau\rightarrow \tau$, $\sigma\rightarrow\sigma+\Delta_{ab}\beta^b+\frac12\tau\Delta_{aa}$, $\beta^b\rightarrow\beta^b+\tau\delta_a^b$. }.
 
 The transformations (\ref{duality}) generate the U-duality group of our system which will be denoted by ${\rm O}(\Gamma_{2,2+n})$ and is isomorphic to ${\rm O}(2,2+n;\mathbb{Z})$. 
 Here $\Gamma_{2,2+n} \simeq \Gamma_{2,2}\oplus \Gamma_n$ with $\Gamma_{2,2}$ the unique four-dimensional self-dual even lattice of signature $(2,2)$ while  $ \Gamma_n$ is a sub-lattice of $H^{2}(X;\mathbb{Z})$
generated by $n$ integer self-dual forms $\chi_a\equiv[\calc_a]$ equipped with the positive definite pairing $\Delta_{ab}$ defined in (\ref{deltab}).  

\section{U-folds: the six-dimensional perspective}
\label{sec:IIBcomp}
 \label{sec:topcond}
 
 The six-dimensional effective theory describing the dynamics of type IIB supergravity on K3 can be obtained by dimensional reduction. First, we split the ten-dimensional space in $M_6\times X$ and we parametrize the  ten-dimensional Einstein frame metric as 
\be\label{six+four}
\d s^2_{\rm E}=e^{2A}\d s^2_{6} + e^{-2A}\d s^2_X
\ee
The warp factor $A$ is taken approximately constant along $X$. After reduction to six-dimensions, 
the ten-dimensional Einstein-Hilbert term reduces to 
\be
  \int\d^{10}x\sqrt{-g_{(10)}}\,R_{(10)}= M^4_{\rm P}  \int\d^6x\sqrt{-g_6}\, R_6+\ldots  \label{10d6d}
\ee
with
  \be\label{MP}
M^4_{\rm P}  \equiv -\frac12\int_X j\wedge j 
\ee
   the K3 volume computed by the metric $\d s^2_X$. We use units where $2\pi \sqrt{\alpha'}=1$ and denote
   by  $M_{\rm P}$   the six-dimensional Planck mass. On the other hand, the effective dynamical volume of the internal space $X$ is given by
 \be
  {\cal V}=-\ft{1}{2   }  \int_X e^{-4A-\phi}\, j\wedge j  
  = \Im\tau\Im\sigma-\frac12\Im\beta\cdot \Im\beta
   \ee
    where  $\Im\beta\cdot \Im\beta\equiv \Delta_{ab}\Im\beta^a\Im\beta^b$. 
   In addition to $\cal V$, the moduli space of metrics on K3 contains other 57 moduli that specify a  hyperk\"ahler structure, i.e.  a choice of a triplet of anti-self-dual two-forms $(\Re\omega,\Im\omega,j)$ inside the space $\Rint^{3,19}$ of two forms on K3.
 The remaining moduli come from the dilaton $\phi$,  the axion $C_0$, one scalar associated with $C_4\in H^4(X;\mathbb{R})$
  and  44 scalars coming from $B,C_2\in H^2(X;\mathbb{R})$, since $\dim H^2(X;\mathbb{R})=22$.  Altogether, these fields parametrize  the moduli space 
\be\label{modulispaceR}
\calm_{\text{IIB on K3}}={\rm O}(\Gamma_{5,21})\backslash { {\rm O}(5,21;\mathbb{R})\over 
 {\rm O}(5;\mathbb{R}) \times {\rm O}(21;\mathbb{R})}
\ee 
The discrete group ${\rm O}(\Gamma_{5,21})\equiv{\rm O}(5,21;\mathbb{Z}$) is the U-duality group of the effective six-dimensional theory.
 A more detailed description of this moduli space is presented in Appendix \ref{app:trunc}.

\subsection{The reduced moduli space}

 The fields  $\tau,\sigma,\beta^a$  characterizing the supersymmetric vacua under study here parametrize the reduced moduli space 
 \be\label{truncmsn}
\calm ={\rm O}(\Gamma_{2,2+n})\backslash { {\rm O}(2,2+n;\mathbb{R})\over
 {\rm O}(2;\mathbb{R}) \times {\rm O}(2+n;\mathbb{R})}
\ee 
In Appendix \ref{app:trunc} we show that the reduction to (\ref{truncmsn}) defines a consistent truncation of the moduli space (\ref{modulispaceR}) of type IIB supergravity on K3. 
 Acting with  the U-duality group ${\rm O}(\Gamma_{2,2+n})$ on the elementary branes discussed 
 in section  \ref{sechol} 
  one can generate the
monodromies associated with a general system of $(p,q)$ 3-, 5- and 7- branes. 
     We will be interested in describing systems in which different branes are contemporarily present. These are characterized by solutions where the holomorphic fields $\tau,\sigma,\beta^a$ are allowed to jump under the U-duality group of the
     effective six-dimensional theory.  This is the six-dimensional analogue of the more familiar F-theory elliptic backgrounds, in which the U-duality group is just SL(2,$\mathbb{Z})$.   
       
       The space (\ref{truncmsn}) has the important property of being  a K\"ahlerian coset manifold. The fact that it is complex is already evident from its   parametrization provided by the fields introduced in  (\ref{sols0}). It is then possible to show that the standard coset metric can be written as a K\"ahler metric, with K\"ahler potential (see Appendix \ref{app:trunc})
\be\label{kpot}
K=-\log\calv
\ee
with 
\be
  {\cal V}=  \Im\tau\Im\sigma-\frac12\Im\beta\cdot \Im\beta
     \label{calv}
   \ee
 It is important to notice that the fields  $\tau,\sigma,\beta^a$
parametrizing (\ref{truncmsn}) must satisfy the conditions  
\be\label{podefcond}
\Im\tau,\Im\sigma,\calv> 0
\ee and therefore
 $K$ is real as expected. 
 $K$ defines a well defined metric  on the orbifolded coset space $\calm$  as well. Moreover, one can easily check that $\calv$ is invariant under $T$, $W_{a}$ and  $R$  in (\ref{duality}), while it transforms as $\calv\rightarrow \calv/|\tau|^2$ under $S$. Correspondingly,
\be
S  :\quad K \rightarrow K-\log\tau-\log\bar\tau
\ee
which is a K\"ahler transformation. Since this happens for the generators (\ref{duality}), the same is true for any element of ${\rm O}(\Gamma_{2,2+n})$. 

 \subsection{Six-dimensional equations  }
 
  We are interested on solutions of this six-dimensional supergravity involving non-trivial backgrounds for the metric and the scalar fields  $\varphi^I=(\tau,\sigma,\beta^a)$ spanning  the K\"ahlerian coset submanifold (\ref{truncmsn}). The setting is completely analogous to the one considered in \cite{cstring}. The relevant terms of the effective action are:
\be\label{effact}
S_{\rm eff}=\frac{M^4_{\rm P}}{2} \int \d^6 x\,\sqrt{-g} \Big[R-2K_{I\bar J}(\varphi)\nabla_M\varphi^I\nabla^M\bar\varphi^{\bar J}\Big]
\ee
where  $K_{I\bar J}\equiv \frac{\del^2 K}{\del\varphi^I\del\bar\varphi^{\bar J}}$
is the K\"ahler metric associated with the K\"ahler potential (\ref{kpot}).
In general, assuming that the complex scalars  $\varphi^I$ depend just on  one complex coordinate $(z,\bar z)$,  for any K\"ahlerian target space the scalars equations of motion reduce to
\be\label{scaleq}
\partial_z\partial_{\bar z}\varphi^I+\Gamma^I{}_{JL}\partial_z\varphi^J\del_{\bar z}\varphi^L=0
\ee
where $\Gamma^I{}_{JK}$ are the Christoffel symbols of the K\"ahler metric, which have crucially only purely holomorphic
or anti-holomorphic indices. 
 
The equations (\ref{scaleq}) are easily solved  by choosing the scalar fields to be holomorphic,  namely
\be\label{holcond}
\delbar\tau=0\, ,\quad \delbar\sigma=0\, ,\quad \delbar\beta^a=0
\ee
We see that this purely six-dimensional description  reproduces the conditions  (\ref{holcond0}) obtained by direct integration of  the ten-dimensional supersymmetry equations.

On the other hand, one can take an ansatz for the six-simensional metric of the form $\d s^2=\d x^\mu\d x_\mu+e^{\rho(z,\bar z)}\d z\d \bar z$. By using once again the K\"ahler structure  defining the effective action (\ref{effact}), the Einstein equations reduce to $\del\delbar(\rho+K)=0$, where $K$ is the K\"ahler potential (\ref{kpot}). Hence, the Einstein equations are solved by the $6$-dimensional metric   
\be
\d s^2_6=\d x^\mu \d x_\mu+ M_P^{-4}\,  {\cal V }\, |h(z)|^2\, \d z \d\bar z \label{g6}
\ee
where $\calv$  is defined in (\ref{calv})   and the constant factor $M_P^{-4}$ is included for matching  the 10-dimensional metric (\ref{10dmetric}).

 We notice that the invariance of the metric (\ref{g6}) under the U-duality group implies that the function $h(z)$ should transform under $S$ in such a way to keep  $\calv |h(z)|^2$ invariant. This requirement is satisfied if we choose 
\be
h(z) =\left( {\chi_D(\tau(z),\sigma(z),\beta^a(z))\over \prod_{i} (z-z_i)  } \right)^{1\over D}   \label{hd}
\ee
 where $\chi_D$ is a generalized ${\rm O}(2,2+n,\Zint)$ modular form of weight $D$, i.e. a 
 function of $\varphi^I=(\tau,\sigma,\beta^a)$  
 which is  invariant under the generators $T,W_a,R$ in (\ref{duality}) and transforms  under $S$ as
\be\label{holmodular}
\chi_D (S\cdot\varphi^I)=\tau^D \, \chi_D (\varphi^I)
\ee  
Finally $z_i$ are simple zeros of  $\chi_D(\tau(z),\sigma(z),\beta^a(z))$, with the denominator $\prod_{i} (z-z_i)  $ in (\ref{hd}) included in order to cancel the zeroes of  $\chi_D(\varphi^I(z))$ in $\Cint$, leaving a no-where vanishing function $h(z)$. 
We will see that the choice (\ref{hd}) is compatible with supersymmetry requirements too.

 
 \subsection{Supersymmetry and topological conditions}
 \label{sec:topcond}
 
 In order to show that the six-dimensional backgrounds described in this section  are supersymmetric, one has to show the existence  of Killing spinors under which the supersymmetric variations of the six-dimensional gravitino and matter fermions vanish.  The supersymmetry conditions follow from those of ${\cal N}=(2,0)$ supergravity  \cite{romans} after reduction
to the truncated moduli space (\ref{truncmsn}). 
This problem is somewhat technical and for this reason its discussion is detailed in Appendix \ref{app:susy}. Here we quote the main results.

As discussed in Appendix \ref{app:susy}, one can in fact write down an ansatz for eight independent Killing spinors (hence providing four-dimensional $\caln$=2 supersymmetry) written just in terms of a single two-dimensional chiral spinor $\eta$ defined on the $z$-plane. The supersymmetry variations of the matter fermions are vanishing once the fields $\varphi^I=(\tau,\sigma,\beta^a)$ are holomorphic, as in (\ref{holcond}). On the other hand,   
  the vanishing of the gravitino variation reduces to the following two-dimensional equations
\be\label{2dsusy6}
D_m \eta\equiv(\nabla_m-\frac{\ii}{2}\calq_{m}  )\eta=0
\ee
where the index $m=1,2$ runs over coordinates of the complex plane.  $\nabla_m$ is the covariant derivative associated with the ordinary spin connection, which must be computed by  using the two-dimensional metric $\calv |h(z)|^2\d z\d \bar z$ appearing in (\ref{g6}). $ \calq_{m}$ are the components of the  ${\rm SO}(2)\sim {\rm U}(1)$-connection:
\be\label{calq}
\calq=\calq_{z}\,\d z+\calq_{\bar z}\,\d \bar z=\Im \Big(\frac{\del K(\varphi(z),\bar\varphi(\bar z))}{\del z}\,\d z\Big)
\ee
 It is easy to see that 
the $\calv$-dependent contribution to the spin connection in (\ref{2dsusy6}) cancels against $\calq_m$ leading to 
the Killing spinor solution
\be\label{globeta0}
\eta=\left(\frac{h(z)}{\bar h(\bar z)}\right)^{\frac14}\eta_0
\ee
with constant $\eta_0$ which satisfies the appropriate projection conditions -- see appendix \ref{app:susy}.


It is well known that codimension-two configurations generically produce deficit angles at large distances \cite{Deser:1983tn} and this puts severe consistency constraints. If we assume to have a configuration in which the holomorphic fields $\varphi^I(z)$ are asymptotically constant for $|z|\rightarrow \infty$, the deficit angle at infinity is given by $\Delta\theta=\int \calr_{(2)}$, where    $\calr_{(2)}$ is the two-dimensional ${\rm SO(2)}\simeq {\rm U}(1)$ curvature. On the other hand, the integrability of (\ref{2dsusy6}) requires that $[D_m,D_n]\eta=0$ and then $\calr_{(2)}=F_\calq$, where $F_\calq =\d\calq$ is the curvature associated with the U(1) connection $\calq$.  Hence, supersymmetry requires the deficit angle to be given by $\Delta\theta=\int \calr_{(2)}=\int F_\calq$, consistently with the results of section 5 of \cite{cstring}. 

In particular, the transverse space closes up to a sphere $\mathbb{P}^1$ when $\Delta\theta=4\pi$. In this case the holomorphic tangent bundle $\calt_{\mathbb{P}^1}$ is isomorphic to  $\calo_{\mathbb{P}^1}(2)$, which is the line bundle whose sections are homogeneous  polynomials of degree-two  in the projective coordinates $[z_0:z_1]$. Indeed, from $\calr_{(2)}=2\pi\,c_1(\calt_{\mathbb{P}^1})$ one gets  $\Delta \theta=2\pi \int_{\mathbb{P}^1}c_1(\calo_{\mathbb{P}^1}(2))=4\pi$ as required.
 On the other hand, the integrability condition $\calr_{(2)}=F_\calq$ implies that the holomorphic line bundle $\call_Q$ associated with the connection $\calq$ is isomorphic to $\calo_{\mathbb{P}^1}(2)$. Noticing that the pull-back of a modular form $\chi_D(\varphi^I)$ of weight $D$  can be regarded as a section of  $\call^D_Q$, we see that $\chi_D(\varphi(z)^I)$  must be given by a homogeneous polynomial of degree $2D$ in the projective coordinates $[z_0:z_1]$. 
 
 This implies in particular that if
  $\chi_D(\varphi^I(z))$ appearing in (\ref{hd}) has $2D$ zeros in the $z$-plane, the plane compactifies to a $\Pint^1$. Consistently the  metric (\ref{g6}) at large  $z$  is  regular as can be seen from the asymptotic behaviour
  \be
\calv \,|h(z)|^2\, \d z \d\bar z \simeq  \calv_0\,|z|^{-4}\, \d z \d\bar z  = \calv_0\,\d w\d\bar w
\ee
with $w=1/z$ the coordinate on the second chart  of $\Pint^1$.

\section{U-fold solutions without 3-form fluxes}
\label{sec:examples1} 
 
    In this and the next section we present some simple examples of U-folds. 
   We start by considering the case $n=0$, in which the K3 space is smooth and there  are no three-form fluxes.  The  restricted moduli space becomes 
   \be\label{truncms22}
\calm_{n=0}={\rm O}(\Gamma_{2,2})\backslash  { {\rm O}(2,2;\mathbb{R})\over  {\rm O}(2;\mathbb{R}) \times {\rm O}(2;\mathbb{R})}
\simeq  \mathbb{Z}_2\backslash \left(  {\rm O}(\Gamma_{1,1})\backslash  { {\rm Sl}(2;\mathbb{R})\over  {\rm U}(1)}  \right)^2
\ee 
where $ \mathbb{Z}_2$ refers to the $R$-duality.  Being the moduli space factorized,  this case can be considered as a doubled elliptic fibration, a double copy of the well known F-theory elliptic geometries  \cite{Vafa:1996xn}, 
where the torus fiber is replaced by a factorized product of two tori with complex structures $\tau$ and $\sigma$ respectively. 
In analogy with the standard F-theory elliptic geometry \cite{sen1}, the solution describes now a system of 
regular D7 and D3 branes with O7 and O3 planes non-perturbatively resolved in terms of $(p,q)$-branes. 
 The interpretation in our setting is completely analogous to the ordinary F-theory case. 
  For this reason we will be rather sketchy. 

Consider first a non-trival $\tau(z)$. 
  The details of the geometry are encoded in an elliptic curve which can be generally written into the form 
   \be
    y^2 =  \prod_{i=1}^3 (x-e_i(z) )=x^3+f_2(z) x+f_3(z)   \label{ellip}
    \ee 
  This solution describes systems of $(p,q)$ 7-branes, related to the elementary D7 branes  via $SL(2,\Zint)$-duality.      In particular for a choice of $f_2$, $f_3$ where  (\ref{ellip}) matches the Seiberg-Witten curve 
  of a ${\cal N}=2$ SU(2) gauge theory with four fundamentals the
  solution describes the non-perturbative resolution of a system of four D7-branes and one O7-plane \cite{sen1} probed by 
  an elementary D3-brane \cite{Banks:1996nj}. 

The axio-dilaton profile can be extracted from the standard elliptic formula 
\be
  {e_{12}\over e_{13} }= {\theta_2^4\over \theta_3^4}(\tau)  \label{zeta1}
 \ee
     relating the harmonic ratio of the roots to the complex structure $\tau$ of the torus. Here $\theta_{s}$ are the genus one
   even theta constants    (see (\ref{genus1theta}) and (\ref{thetastand}) for the definition). 
     The positions of D7-branes correspond
     to points in the $z$-plane where  $e_1\to e_2$, i.e.  $\tau\to \ii \infty$. Going
     around this point the axio-dilaton field undergoes the 
     monodromy $\tau\to \tau+1$. 
     Finally the O7 plane corresponds to a
     pair of degeneration points with overall monodromy $\tau\to \tau-4$. The effects of instantons resolve this plane into a 
   pairs of $(p,q)$-branes which locally look like D7-branes (in a given SL(2,$\Zint$) frame). Encircling the two
   $(p,q)$ 7-branes one finds the monodromy  reproducing the O7-plane charge \cite{sen1}.

    The story for the $\sigma(z)$ field follows {\it mutatis mutandis} that of $\tau$. Again the details of the geometry 
  are encoded in the elliptic data  
       \be
  {\tilde e_{12}\over \tilde e_{13} }= {\theta_2^4\over \theta_3^4}(\sigma)   \label{zeta2} 
 \ee   
 with $\tilde{e}_i$ the roots of an elliptic curve of type (\ref{ellip}). 
 The fibration describes now the background of $(p,q)$ 3-branes, related to elementary D3-branes via ${\rm SL(2;\mathbb{Z})}$ duality. The elliptic fibration now corresponds to the Seiberg-Witten curve describing the  dynamics of the gauge theories in D7 brane probes of the D3 brane geometry after reduction to four-dimensions. 
    
  Summarizing, the U-fold solution with no three-form fluxes is specified by the choice of two  elliptic curves  fibered over $\Cint$ with punctures  signaling the presence of 3 and 7-branes.  Notice that in this case there is a natural candidate for the holomorphic function $h(z)$ entering the metric of  the solution, which follows from the doubling of the solution of \cite{cstring}. This is given by
      \be
h(z)=\frac{\eta(\tau(z))^2}{\prod_{i} (z-u_i)^{1\over 12}}
 \frac{\eta(\sigma(z))^2}{\prod_{j} (z-v_j)^{1\over 12}}   \label{heta}
\ee
where $u_i$ and $v_j$ are the points where the elliptic fibrations defining $\tau$ and $\sigma$ respectively degenerate. 
These are the points at which the discriminant $\Delta(z)=4f^3_2(z)+27f^2_3(z)$ of one of the two elliptic curves vanishes and they signal in general the presence of  $(p,q)$-branes.  This  is consistent with the general form of $h(z)$ given in (\ref{hd}) with $D=12$ and 
   $\chi_{12}(\tau,\sigma)=\eta(\tau)^{24}\eta(\sigma)^{24}$.  As explained in section \ref{sec:topcond}, 
   for  fibrations chosen such that there are 24 degeneration  points in total, the complex plane compactifies to the sphere $\mathbb{P}^1$.

    %

\section{U-folds from hyperelliptic fibrations}
\label{sec:examples2}

In this section we discuss in some more details the case $n=1$, i.e.\ the case in which the K3 develops a local $\mathbb{C}^2/\mathbb{Z}_2$ singularity with a single exceptional cycle  $\calc_1\equiv \calc$. An analogous discussion for the case with $n>1$ exceptional cycles is left to the future.

The solutions in the case $n=1$ involve three active scalar fields $(\tau,\sigma,\beta)$, where $\beta\equiv \beta^1$ according to the general notation used in the previous sections. These scalars  follow from the reduction of the axio-dilaton field, the warp factor, the RR four-form and the NSNS and RR two-form
potentials along  $ \calc$. 
 The three complex scalars span the coset
 \be\label{truncms}
\calm={\rm O}(\Gamma_{2,3})\backslash { {\rm O}(2,3;\mathbb{R})\over  {\rm O}(2;\mathbb{R}) \times {\rm O}(3;\mathbb{R})}
\ee 
  The exceptional cycle $\calc$ sitting at the singularity is a two-sphere with self-intersection $\calc\cdot\calc=-2$ and then $\Delta_{11}=2$. Hence, in this case,  the generators of the U-duality group (\ref{duality}) reduce to
\be\label{duality2}
\begin{aligned}
&T:\quad\tau\rightarrow \tau+1 \qquad ~~~~~
R:  \quad\tau\leftrightarrow \sigma~~~~~  \qquad W:\quad  \beta\rightarrow \beta+1\\
&S:\quad\tau\rightarrow -\frac{1}{\tau} \qquad \qquad \sigma\rightarrow \sigma-\frac{1}{\tau}\,\beta^2    \qquad
  \beta\rightarrow \frac{1}{\tau}\beta  
\end{aligned}
\ee
 We observe that the U-duality group ${\rm O}(\Gamma_{2,3})$ generated by (\ref{duality2}) is isomorphic to the modular group   Sp$(4,\Zint)$ 
of a genus two hyperelliptic Riemann surface.
Moreover, if we organize the three complex scalars into a $2\times 2$ matrix
\be
{\bf \Omega}=\left(\begin{array}{cc} \tau & \beta \\ 
\beta & \sigma \end{array}\right)
\ee  
one can see that  ${\bf \Omega}$ transforms under the U-duality transformations  (\ref{duality2}) as 
 the period  matrix  of a genus two Riemann surface (see appendix \ref{sgtwo})
\be
{\bf\Omega}\rightarrow (A{\bf\Omega}+B)(C{\bf\Omega}+D)^{-1}
\ee 
with $A,B,C,D$ $2\times 2$ matrices defining a matrix of Sp$(4,\Zint)$ via\footnote{
Explictly
$
{\tiny S=\left(\begin{array}{cccc}
 0  & 0 & -1 & 0      \\
 0 &  1 & 0 & 0 \\
 1 &  0 & 0 & 0 \\
 0 & 0 & 0 & 1 
\end{array}\right)}\, , \,
{\tiny T=\left(\begin{array}{cccc}
 1  & 0 & 1 & 0      \\
 0 &  1 & 0 & 0 \\
 0 &  0 & 1 & 0 \\
 0 & 0 & 0 & 1 
\end{array}\right)}\, , \, {\tiny R=\left(\begin{array}{cccc}
 0  & 1 & 0 & 0      \\
 1 &  0 & 0 & 0 \\
 0 &  0 & 0 & 1 \\
 0 &  0 & 1 & 0 
\end{array}\right)}\, , \,{\tiny W=\left(\begin{array}{cccc}
 1  & 0 & 0 & 1      \\
 0 &  1 & 1 & 0 \\
 0 &  0 & 1 & 0 \\
 0 &  0 & 0 & 1 
\end{array}\right)}$
}
 \be\label{spmatrix}
 M=
\left(
\begin{array}{cc}
 A  & B      \\
 C &   D   \\ 
\end{array}
\right)
  \qquad 
 M  \left(
\begin{array}{cc}
 0  & \bbone     \\
 -\bbone &   0   \\ 
\end{array}
\right) M^T=  \left(
\begin{array}{cc}
 0  & \bbone      \\
 -\bbone &   0   \\ 
\end{array}
\right)
  \ee
 Furthermore,  the quantity $\calv$ defined in (\ref{calv})
 reduces in this case to
\be
\calv=\Im\tau\Im\sigma-(\Im\beta)^2\equiv \det\Im{\bf \Omega}
\ee
Hence, the consistency condition $\calv>0$ translates into the condition that $\Im{\bf \Omega}$ is a positive definite matrix, as required for ${\bf \Omega}$ being the period matrix of a Riemann surface. 

 The identification of ${\bf \Omega}$ with the period matrix of a genus two Riemann surface suggests  that 
 the U-fold solution can be viewed as a holomorphic  fibration  of a genus two Riemann surface  over the complex plane $\Cint$. The period matrix ${\bf \Omega}(z)$  describes the variations of scalar fields over the complex plane $\Cint$.  U-duality holonomies around brane locations are encoded in the  non-trivial modular group transformations that the cycles of the Riemann surface undergo around a point where the  fiber degenerates. 
 
  The geometry of genus two fibrations over a complex plane has been extensively studied  in the
 mathematical literature and one can resort to this powerful apparatus to explore the physics of U-folds in this sector.
 Here we will not attempt an analysis of the general case but rather we focus  on some explicit choices of fibrations
 illustrating  few relevant features of the general solution. We start by describing the geometry of the genus two curve,
 the period matrix ${\bf \Omega}(z)$, its degenerations, holonomies and brane interpretation.   
  
 \subsection{The genus two fibration}
 
We start by describing the geometry of the hyperelliptic fibration. We refer the reader
to  Appendix  \ref{sgtwo} for further details. 
  A Riemann surface of genus two can be 
  always described by a hyperelliptic curve (a sextic or a quintic)
   \be
   y^2 =\prod_{i=1}^6 (x-e_i(z) ) =x^6+f_2(z) x^4+ f_3(z) x^3+\ldots+f_6(z)   \label{sucurve0}
   \ee
    At each point $z$ equation (\ref{sucurve0}) specifies a genus two curve. 
      The period matrix ${\bf \Omega}(z)$ of the genus two fiber at $z$ is computed by integrals around the non-trivial cycles in the complex $x$-plane with three cuts pairing the six 
      roots $e_i$ (see Appendix \ref{AppCycles} for details).
 
 Alternatively, the hyperelliptic curve can be written directly in terms of the theta functions  of
 the genus two Riemann surface $\theta[^a_b ] =\theta[^a_b] (0|{\bf \Omega})$ with half-characteristics $[^a_b]$.
 Indeed after using SL$(2,\Rint)$ invariance to map, let us say,  points $e_1,e_3,e_5$
to $0,1,\infty$ the curve can be brought to the quintic form
   \be\label{xipar}
 y^2=x(x-1)(x-\xi_2)(x-\xi_4)(x-\xi_6)
 \ee
with
 \be
 \begin{aligned}
 \xi_2({\bf\Omega}) &=&{ e_{21} e_{35}\over e_{25} e_{31} } ={ \theta[^{11}_{11}]^2({\bf\Omega})\, \theta[^{10}_{00}]^2 ({\bf\Omega})\over \, \theta[^{01}_{00}]^2({\bf\Omega})\, \theta[^{00}_{11}]^2({\bf\Omega})}  \\
  \xi_4({\bf\Omega}) &=& { e_{41} e_{35}\over e_{45} e_{31} } ={ \theta[^{10}_{00}]^2({\bf\Omega})\, \theta[^{00}_{10}]^2({\bf\Omega})\over \theta[^{00}_{11}]^2({\bf\Omega})\, \theta[^{10}_{01}]^2({\bf\Omega})}  \\
    \xi_6({\bf\Omega}) &=& { e_{61} e_{35}\over e_{65} e_{31} }  ={ \theta[^{00}_{10}]^2({\bf\Omega})\,\theta[^{11}_{11}]^2({\bf\Omega})\over \theta[^{10}_{01}]^2({\bf\Omega})\, \theta[^{01}_{00}]^2 ({\bf\Omega})}    \label{xi2340}
    \end{aligned}
 \ee
    The genus two surface degenerates whenever two roots $e_i$ collide signaling for the presence of a brane.    
 In general,  a degeneration shows up in the vanishing of the discriminant  of the curve, that we denoted by
 $I_{10}$ and which is defined by 
    \be
     I_{10} = \prod_{1\leq i<j \leq 6} e_{ij}^2  
   \ee
    For $I_{10} \neq 0$ the Riemann 
surface is smooth. At a point  $z=z_0$ where the discriminant vanishes   
 the genus two curve   degenerates.  Going around $z_0$, the period matrix  ${\bf \Omega}(z)$ undergoes non-trivial monodromies. Given a hyperelliptic fibration,  the brane content of the system is specified by these monodromies and the full non-perturbative dynamics is coded in the details of the fibration. 

\begin{figure}[!t]
\centering
\includegraphics[scale=.25]{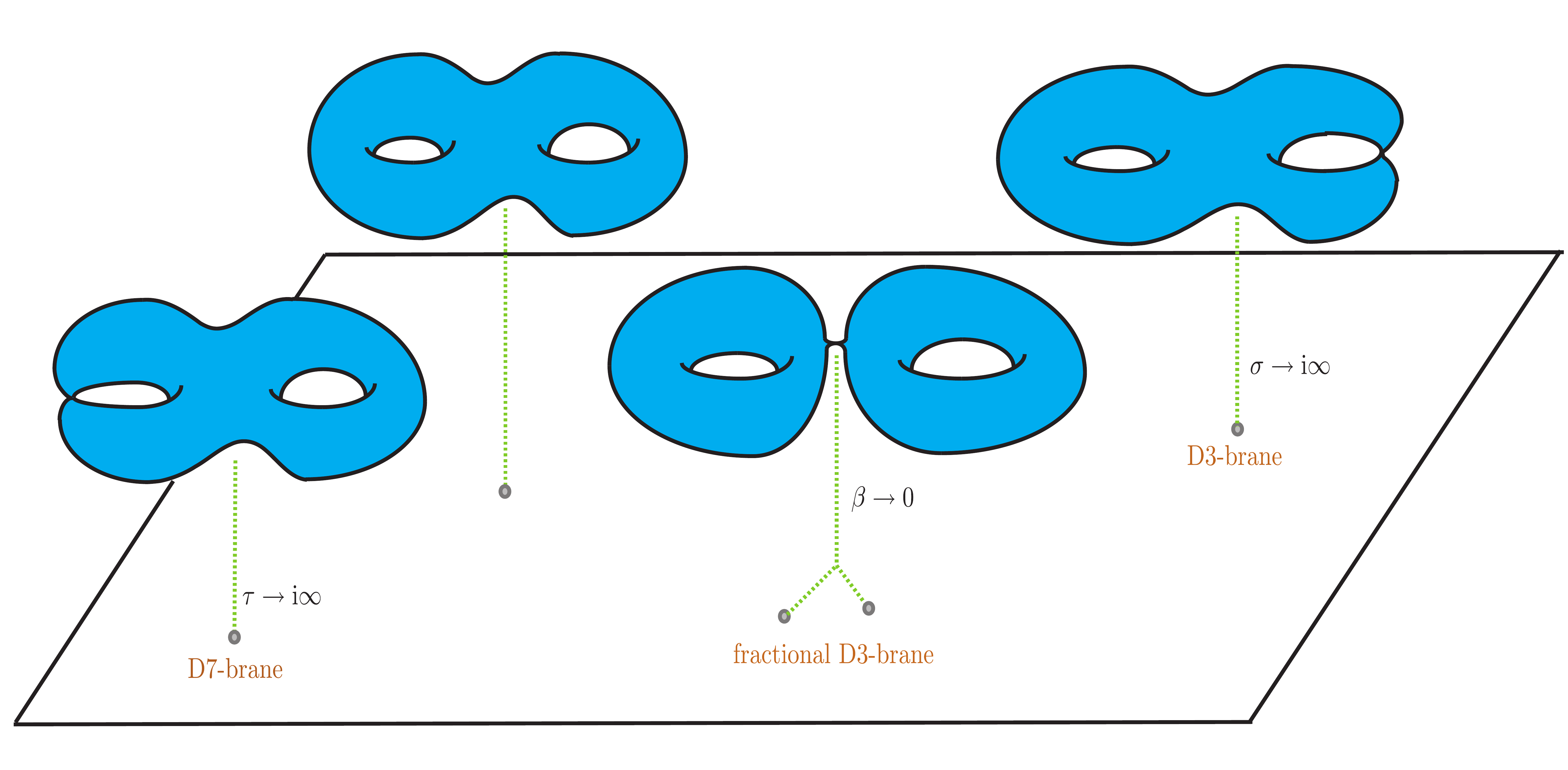}
\caption{\small The basic degenerations of the genus two fiber.}
\label{fig:deg}
\end{figure}

    There are various basic ways in which a genus two Riemann surface can degenerate, see figure \ref{fig:deg}. First, when one of the two handles is pinched, the Riemann surface degenerates to a torus  with a double point. This happens if $\tau\rightarrow \ii\infty$ or  $\sigma\rightarrow \ii\infty$, signaling the presence of 7- and 3-branes respectively.  
 A genus two surface can also degenerates into two genus one surfaces when $\beta\to 0$. This degeneration will show up, for example, in the solution representing flux-dissolved fractional 3-branes in section \ref{sec:fracD3}. On the other hand, a localized D5 brane charge would be signalled by a degeneration $\beta\to \infty$ which, however, can never come alone since $ \Im\tau\Im\sigma-(\Im\beta)^2>0$ for a genus two Riemann surface.
%

Finally, let us recall that in order to completely specify the background, one has to specify the modular form 
$\chi_D({\bf \Omega})$ entering in the metric.  
  In analogy to F-theory elliptic solutions we expect $\chi_D({\bf \Omega})$ to vanish at the positions of 3- and 7-branes. This suggests that $\chi_D({\bf \Omega})$ is not only a modular form but a cusp form. The ring of cusps
forms is generated by the three functions  $\chi_{10}({\bf \Omega})$,  $\chi_{12}({\bf \Omega})$ and $\chi_{35}({\bf \Omega})$
whose definition is provided in Appendix \ref{sgtwo}. It is not clear to us whether any choice  of the cusp form defines an admissible solution or if  there is a privileged one. We postpone this interesting question to future investigations.

   \subsection{Fractional D3-branes}
   \label{sec:fracD3}


In this section we describe the solution associated to a system of fractional D3 branes at a $\mathbb{C}^2/\mathbb{Z}_2$ singularity. This corresponds to the limit of large volume $\sigma \to \ii \infty$ and weak coupling $\tau \to \ii \infty$ of  the hyperelliptic fibration. We will first show how the solution can be entirely reconstructed from the knowledge of its
monodromies (the brane content). The results will be then matched with those coming from the hyperelliptic description.  

A system of fractional D3 branes at $\mathbb{C}^2/\mathbb{Z}_2$ is characterized by two integers $N_0,N_1$ counting
the number of fractional branes of each type. The brane system realizes a ${\cal N}=2$ quiver gauge theory with gauge
group $U(N_0)\times U(N_1)$ and bifundamental matter. On the other hand fractional branes are source for the twisted 
field $\beta$ describing the NS-NS/R-R two-form fluxes along the exceptional cycle $\calc$ of the singularity. 
Since fractional D3-branes can be thought as D5-branes wrapping the exceptional cycle $\calc$ 
with opposite orientation, one expects a monodromy  $\beta\rightarrow\beta+N_0-N_1$ for a turn around the location
of the brane system in the $\Cint$-plane.
 One would then be tempted to locally  describe them by $\beta\simeq \pm\frac{1}{2\pi\ii}\log(z-z_{\rm D5})$. However, this creates a problem. Around the D5-brane, $\tau$ and $\sigma$ are expected to be approximatively constant and therefore  at some point the condition $\calv=\Im\tau\Im\sigma-(\Im\beta)^2 > 0$ would be violated, which would be inconsistent with the formulation based on the hyperelliptic fibration. 
 
In order to address this problem,  let us consider a local description of the system of fractional D3-branes. Practically, this can be obtained by taking
  a very large K3 volume. According to (\ref{MP}), this corresponds to  the limit of large Planck mass $M_{\rm P}\gg 1$, where then the six dimensional gravity effectively decouples.  Furthermore,  we take $\Im\sigma$, $\Im\tau$ large in order to keep 
  the coupling of the string small and the combination $M_P^{-4} \,\calv$ appearing in the six-dimensional metric
  $(\ref{g6})$ finite.  
  Under these conditions, the only allowed  U-duality transformations are those leaving $\tau$ and $\sigma$ invariant. These
  transformations are
  generated by the two elements
 \be
W : \quad \beta\to \beta+1   \qquad~~~~~~~~~~ S^2:\quad  \beta \to -\beta     
 \ee
A holomorphic function $\beta(z)$   with holonomies only of this kind can be written as
  \be
\beta(z) =\ft{1}{4\pi \ii} \ln \left[ { P_0(z)+\sqrt{P_0^2(z)- P_1^2(z) }}\over{ P_0(z)-\sqrt{P_0^2(z)- P_1^2(z) } }\right] \label{betapq}
     \ee
 with $P_0(z), P_1(z)$ polynomials in $z$ of order $N_0$ and $N_1$ respectively. 
 The function $\beta(z)$ displays indeed the following monodromies
 \bea
z_0=\infty : &&\qquad  \beta \rightarrow \beta+N_0-N_1 \nn\\
\{ z_0~; ~P_1(z_0)=0\}: &&   \qquad \beta \rightarrow \beta-1\nn\\
\{ z_0~; ~P_0(z)^2=P_1(z_0)^2\}: && \qquad  \beta \rightarrow -\beta
 \eea
  We stress that the solution (\ref{betapq}) is valid in the patch of $\Cint$ 
 where $(\Im\beta)^2<\Im\sigma\, \Im \tau$.  
 Hence, for $z$ large enough or near the zeros of $P_1(z)$, the local approximation breaks down and  one should resort to the complete hyperelliptic description. In particular for $N_1=0$ (the pure gauge theory) $\beta(z)$ given in (\ref{betapq}) is locally completely finite, with no logarithmic  singularities at finite $z$, but  is  rather characterized by $2N$ points with monodromy $\beta\rightarrow -\beta$.
 In particular, there are no localized D5-brane sources in this case but  a purely flux solution.\footnote{Notice that, in our hyperelliptic formulation, the zeroes of $\beta$ correspond to the points where the fiber degenerates into to two genus one surfaces.} 

  Formula (\ref{betapq}) agrees with that one proposed in \cite{cremonesi}, motivated by the M-theory lift of the D-brane system, and recently derived in \cite{lerdaetnoi} from a direct string computation \footnote{ Our field $\beta$ is related to the fields $\gamma$ and $t$ in
  \cite{cremonesi,lerdaetnoi} via the identifications  $\gamma=\ft12 t = \beta_1=-\Delta_{11}\beta^1= -2\beta$.}.  We refer to  these references for a more detailed discussion on the physics and holographic interpretation of this solution.
  

\subsubsection{The hyperelliptic description} 

Let us now show how the previous results fit (and generalize) into the general hyperelliptic description of U-folds. We consider a hyperelliptic fibration in the limit $\Im\sigma\rightarrow\infty$, while keeping
$\tau$ approximately constant but {\em finite}. This case covers a class of backgrounds which are dual to one of the M-theory `elliptic models' of  \cite{Witten:1997sc}, see also \cite{Ennes:1999fb,Petrini:2001fk}. These models can be used to extract $\beta(z)$ from the dual M5 geometry, as in \cite{cremonesi}.  

Using the parametrization (\ref{xipar}), the genus two curve is specified by the three harmonic ratios $\xi_2,\xi_4,\xi_6$. Their dependence on the period matrix ${\bf \Omega}$  can be expressed in terms of the theta functions by means of the relations (\ref{xi2340}). One finds that in the degenerate limit $\sigma\to\ii \infty$ the three harmonic ratios become (see Appendix \ref{spinching})
 \be
 \xi_2 =-{\theta_2^2(\tau)\over \theta_4^2(\tau)}{\theta_1^2(\ft{\beta}{2}|\tau)\over \theta_3^2(\ft{\beta}{2}|\tau)} \qquad    \xi_4=1 \qquad \xi_6=-{\theta_4^2(\tau)\over \theta_2^2(\tau)}{\theta_1^2(\ft{\beta}{2}|\tau)\over \theta_3^2(\ft{\beta}{2}|\tau)} \qquad~~~\text{(for $\sigma\rightarrow \ii\infty$)} \label{oneh2}
 \ee
  with the genus one theta functions given by (\ref{thetastand}).  
 The fact that $\xi_4=1$ means that the genus two curve has degenerated to a two-torus, as already mentioned.
 The requirement that $\tau$ remains constant over the complex plane
  translates into the condition that the ratio ${\xi_2(z) \over \xi_6(z)}$ is constant over $\Cint$. 
 Inverting the last relation in (\ref{oneh2}) one can then find $\beta(z)$ in terms of the harmonic ratio $\xi_{6}(z)$.
  
If we also perform the limit $\tau\to\ii \infty$ the three harmonic ratios further degenerate (see Appendix \ref{twopinch} for
 details)
 \be
 \xi_2 =0   \qquad \xi_4=1  \qquad  \xi_6=-\sin^2\left(\frac{\pi\beta}{2}\right)\label{oneh0}\qquad~~~\text{(for $\sigma,\tau\rightarrow \ii\infty$)}
 \ee
 This implies that the combination
 \be
 f(z)=\cos 2\pi \beta(z)=2(2\xi_6(z)+1)^2-1  \label{ff}
 \ee
 is a well defined function on the $\Cint$-plane (invariant under $\beta\to -\beta$ or $\beta \to \beta+1$).
 Inverting (\ref{ff}) one finds
  \be
  \beta_+(z)=-\beta_-(z)=\ft{1}{4\pi \ii} \ln \left[ { f(z)+\sqrt{f(z)^2- 1 } \over f(z)-\sqrt{f(z)^2- 1}}\right] \label{betapq0}
  \ee
 We immediately see that the result (\ref{betapq}) is obtained by setting $ \beta(z)=\beta_+(z)$ and $f(z)=P_0(z)/P_1(z)$. 

\subsection{An example with fractional and regular branes}

 Now we consider a case with both regular and fractional D-branes. A strategy for providing a concrete example, followed for instance in \cite{sen1} for the case of F-theory, is to exploit the experience coming from Seiberg-Witten (SW) theory. Hence, we are led to consider the hyperelliptic curve describing the ${\cal N}=2$ gauge theory with gauge group  SU(3) and six  fundamental hypermultiplets. For simplicity we take the SU(3) theory in the so called special vacua \cite{Argyres:1999ty}, parametrized 
 by three parameters, the cubic gauge invariant $z={\rm tr} \Phi^3$, a mass $m$ and the gauge coupling $g$.  For this
 choice the hyperelliptic curve takes the simple form
 \be
 y^2=\prod_{i=1}^6 (x-e_i(z)  )=(x^3-z)^2-g^2 (x^3-m^3)^2 \label{su3}
 \ee
  The six branch points are given by
\be
  e_{2k}(z)=w^{k-1} \left({ z+ g\, m^3\over 1+ g} \right)^{1\over 3}
  \qquad e_{2k-1}(z)=w^{k-1}  \left({ z- g\, m^3\over 1- g} \right)^{1\over 3}
  \ee
with  $ w=e^{2\pi \ii \over 3}$ and $k=1,2,3$. Let us consider first the asymptotic geometry.  
At $z$ infinity one finds
\be
  e_{2k}= {w^{k-1} z^{1\over 3} \over (1+g)^{1/3}}  \qquad    e_{2k-1}= {w^{k-1} z^{1\over 3} \over (1-g)^{1/3}}\quad~~~~~~~~~~\text{(for $z \rightarrow \infty$ )}
\ee
 Plugging this into (\ref{xi2340}), one finds that the harmonic ratios $\xi_{2,4,6}(z)$, and therefore ${\bf \Omega}(z)$ go to a constant
 and finite value for $z\rightarrow \infty$.  
 Furthermore, in  the limit $g\to 0 $, $e_{12},e_{56}\sim g\, z^{1/3}$, and all others $e_{ij}$ go like $z^{1/3}$.  For the harmonic ratios $\xi_{2,4,6}(z)$  we have
 \be
 \xi_2 \sim  g \qquad~~\xi_4 \sim 1\qquad~~ \xi_6\sim g^{-1}   \quad~~~~~~~~~~\text{(for $z, g^{-1} \rightarrow \infty$)} \label{xig}
 \ee  
  To see what this implies for ${\bf \Omega}$, we consider the expansion for  very large  $\Im\tau,\Im\sigma$ of the theta functions in the right hand side of (\ref{xi2340}). By using (\ref{qqy}) and (\ref{thetaexp}), one finds  
 \be
 \xi_2 \sim e^{\pi\ii(\tau-\beta)} \qquad 
 \xi_4 \sim e^{\pi\ii(\tau-\sigma)}   \qquad
 \xi_6 \sim   e^{-\pi\ii\beta}  
\quad~~~~~ \text{(for $\Im\tau,\Im\sigma\gg 1$)}
  \label{xizm2}
 \ee
 Comparing with (\ref{xig}) we conclude that asymptotically for  $z\gg 1$ and in the limit $g\ll 1$ we have the following limiting values of $\tau,\sigma,\beta$:
 \be
 \tau_0\simeq \sigma_0 \simeq \frac12 \beta_0\simeq \frac{1}{2\pi\ii}\log g^4   
  \ee
  In other words the weak coupling of the auxiliary gauge theory  corresponds to the limit where the imaginary parts of all entries 
  of the period matrix are very large, which indeed defines the weak coupling of the string theory description.
  
  Now let us consider the points where the genus two fiber degenerates. From the point of view of the auxiliary gauge theory they correspond to points in the Coulomb branch of the moduli
 space where BPS (in general dyonic) states become massless.
The   discriminant of the curve is given by
 \be
 I_{10}=\prod_{i<j} e_{ij}^2={(6g)^6\over (1-g^2)^8} \,(z-m^3)^6(z^2-g^2 m^6)^2\label{i10} 
 \ee
 The zeros of $I_{10}$ signal the collision of some of the branch points.  
    For the present case, there are three degeneration points 
 (punctures) in the $z$-plane
  \bea
  z=m^3  && \qquad   e_{2k}=e_{2k-1}=w^{k-1} \, m      \nn\\
  z=g\, m^3  && \qquad   e_{1}=e_3=e_5=0  \nn\\
  z=-g\, m^3  && \qquad   e_{2}=e_4=e_6=0  
  \eea  
  for $k=1,2,3$.  Going around these points the period matrix ${\bf \Omega}(z)$ 
  undergoes non-trivial monodromies which characterize the brane content. 
To compute them, we use again the representation (\ref{xi2340}) 
 to compute the period matrix  ${\bf \Omega}(z)$  in the
 nearby of the singularities.
 
 Let us first consider  the geometry near the degeneration point $z= m^3$. For  $z\simeq m^3$,  $e_{2k-1,2k} \sim (z-m^3)$ implying
 \be
   \xi_{2} \sim (z-m^3) \quad , \quad 
 \xi_4\sim 1 \quad , \quad  \xi_6\sim (z-m^3)^{-1}  
 \ee
 Following the same arguments as before we conclude that close to $z= m^3$ we have
 \be
\tau,\sigma \sim \frac{1}{2\pi\ii} \log(z-m^3)^4   
  \qquad   \beta \sim   \frac{1}{2\pi\ii} \log(z-m^3)^2 \quad~~~~~~~~\text{(for $z\simeq m^3$)}
 \ee
 We conclude that all the three entries of the period matrix behave logarithmically as $z\simeq m^3$, leading to the monodromies
\be
\tau \to \tau+4 \qquad~~    \sigma\to \sigma+4 \qquad~~ \beta\to \beta+2 \quad~~~~~~~\text{(around $z= m^3$)}
\ee
 This indicates the presence of 4 D7 branes, 4 D3 branes and 2 fractional D3 branes at $z=m^3$. Notice that in this case the fractional D3-branes are superimposed on D7-branes and D3-branes. This allows, in contrast with the solution
 discused in section  \ref{sec:fracD3},  for the presence of a fractional brane  at a finite distance since $\Im \beta$ can diverge without  violating the consistency condition $\calv >0$.   
 
The monodromies around $z=\pm g m^3$ can be studied  in a similar way. Since at infinity there is no net monodromy, the total monodromy around these two points should be such that   
 it compensates for those coming from the D-branes at $z=m^3$, i.e. 
 \be\label{Omon}
 \qquad \tau \to \tau-4 \qquad~~    \sigma\to \sigma-4 \qquad~~ \beta\to \beta-2 
\ee
We notice that in the perturbative limit $g\ll 1$ the  two degeneration points $z=\pm g m^3$ become very close and separated by a distance $\Delta z\sim m^3 e^{-\frac\pi2\Im\tau_0}\sim m^3 e^{-\frac\pi2\Im\sigma_0}$, which is exponentially suppressed in this weak coupling limit. 
Hence, by analogy with the F-theory case discussed in \cite{sen1,sen2}, it is natural to regard the solution around the degeneration points $z=\pm g m^3$ as a system of mutually non-local branes which provide the non-perturbative resolution of a system of O-planes, with total charges given by the monodromies (\ref{Omon}).   
This is the analogue of the resolution of O7 planes into a pair of $(p,q)$ 7-branes  \cite{sen1},
 which has been explicitly shown to derive from  ED(-1) non-perturbative corrections in \cite{Billo:2011uc}. In the present case we expect that not only ED(-1), but also  fractional ED(-1) and ED3  instantons conspire to resolve the  composite O-planes at $z=0$.
 
 In Appendix \ref{AppPerios} we present an alternative derivation of the monodromies from a direct 
 evaluation of ${\bf \Omega}(z)$ from the period integrals. 



\vspace{2cm}

\centerline{\large\bf Acknowledgments}

\vspace{0.5cm}

\noindent We would like to thank M.~Bianchi, M.~Bill\'o, G.~Bonelli, S.~Cremonesi, F.~Fucito, H.~Samtleben and A.~Uranga for useful discussions.  L.M. would like to thank the Physics Department of Universit\`a di Parma for kind hospitality during the course of this work.
This work is partially supported by the ERC Advanced Grant n.226455 ``Superfields", by the Italian MIUR-PRIN contract 20075ATT78,
by the NATO grant PST.CLG.978785.

\vspace{2cm}

\newpage

\begin{appendix}

\section{Ten-dimensional supersymmetric solutions}
\label{app:10d}
In this Appendix we re-derive the supersymmetry  conditions  presented in  \cite{GKP,GP}, which describe general flux F-theory vacua,
hence characterized by seven-branes, regular or fractional D3-branes as well as compatible bulk fluxes.
These conditions apply to the local ten-dimensional description of our vacua configurations, which admit a global description only at the level of the effective six-dimensional theory after non-trivial U-duality holonomies   are allowed. 
Our aim is to make the paper self-consistent, to clarify the relation between the six- and  ten-dimensional description,
 and facilitate the potential application of our results and
 their comparison with others. Here we use the generalized geometry framework, in which 
 the supersymmetry conditions can be expressed in a compact geometrical form \cite{gmpt} and acquire a clear interpretation from the viewpoint of  D-brane physics \cite{luca1,luca2}.

Take a general type II background preserving four-dimensional Poincar\'e invariance. The ten-dimensional space-time splits as $\mathbb{R}^{1,3}\times Y$ and   the string-frame metric can be written as
\be
{\rm d}s^2_{\rm st}=e^{2D}\d x^\mu\d x_\mu+\d s^2_6
\ee
Four-dimensional $\caln=1$ supersymmetry requires the existence of type II Killing  Weyl spinor $\epsilon=\epsilon_1+\ii\epsilon_2$, which in our case takes the form $\epsilon=\zeta\otimes  \eta$, where $\zeta$ is a constant chiral spinor on $\mathbb{R}^{1,3}$ and $\eta$ is a chiral spinor on $Y$.
We can use the internal spinor $\eta$ to construct the following forms on $Y$:
\be
\Omega_{\rm st}=\frac{1}{3!|a|^2}\,\eta^T\gamma_{mnp}\eta\, \d y^{m}\wedge\d y^n\wedge \d y^p\, ,\qquad  J_{\rm st}=\frac{\ii}{2|a|^2}\,\eta^\dagger\gamma_{mn}\eta\, \d y^{m}\wedge\d y^n
\ee
where $|a|^2\equiv \eta^\dagger\eta$ and $y^m$ are some coordinates on $Y$. The normalization is  taken such that
\be
\ft{1}{3!} J_{\rm st}\wedge J_{\rm st}\wedge J_{\rm st}=-\ft{\ii}{8} \Omega_{\rm st} \wedge \bar \Omega_{\rm st} ={\rm dvol}_6  \label{norm}
\ee
In turn, the information in $J_{\rm st}$ is equivalently encoded in the polyform (alias pure spinor, in generalized geometry language)
\be
\Psi\equiv e^{\ii J_{\rm st} }
\ee 
In \cite{gmpt} it is was proved that the ordinary Killing spinor conditions are equivalent to imposing the following equations on $Y$:\footnote{
In an oriented vielbein $e^a$, the Hodge star is defined by
$ *_d  \omega_p = \frac{1}{p!(d-p)!}\epsilon_{a_1\ldots a_d}\,\omega^{a_{d+1-p}\ldots a_d} \, 
 e^{a_1}\wedge\ldots\wedge e^{a_{d-p}}$, where $\epsilon_{12\ldots d}=1$.}
\bseq\label{susycond}
\begin{align}
\d_H(e^{3D-\phi} \Omega_{\rm st} )&=0 \label{susycond1}\\
\d_H(e^{2D-\phi}\Im \Psi)&=0\label{susycond2}\\
\d_H(e^{4D-\phi}\Re\Psi)&=e^{4D}*_6\lambda(F)\label{susycond3}
\end{align}
\eseq
 where $H$ and $F=F_1+F_3+F_5$ are the field strengths along $Y$ and
\bea
&& \d_H=\d+H\wedge \qquad ~~~~~~~~\lambda(F)=F_1-F_3+F_5
\eea
 Away from localized sources, the fields $H$ and $F$ satisfy the Bianchi identities $\d H=0$ and $\d_H F=0$ which can be locally solved by setting  $H=\d B$ and $F=\d_H C$ with $C=C_0+C_2+C_4$.

By defining
\be
 \Omega\equiv e^{3D-\phi}\Omega_{\rm st} \, ,  \qquad     J\equiv e^{2D-\phi}  J_{\rm st}
\ee
the first two supersymmetry equations (\ref{susycond1},\ref{susycond2}) can be written as
 \be
 0=\d\Omega= \d J = \Omega\wedge H= H\wedge J    \label{domega}
 \ee
 which are equivalent to requiring that the $(\Omega,J)$ define integrable complex and K\"ahler structures respectively. Furthermore,  the 
 condition $J\wedge H=0$ imply that $H$ is primitive. 
 A solution to  (\ref{domega}) is given by taking  
\be
\Omega=h(z)\d z\wedge \omega\, , \quad J=j-\frac{\ii}{2} e^{-\phi} |h(z)|^2 \d z\wedge \d\bar z\, 
\ee
where $\omega$ and $j$ are the anti-self dual 
\footnote{\label{foot:hyper}In a local oriented vielbein $e^a$,  we can write $\omega=(e^1+\ii e^3)\wedge (e^2+\ii e^4)$ and $j=e^{1}\wedge e^3+e^2\wedge e^4 $.  In these conventions ${\rm vol}_4=-\frac12 j\wedge j=-\frac14\omega\wedge \bar\omega$, $*_{X}j=-j$ and $*_{X}\omega=-\omega$.}
  closed two-forms on $X$ for $Y=\mathbb{C}\times X$.  In addition $B$ is taken self-dual to ensure $H\wedge J=0$.  
The forms $J$ and $\Omega$ define the Kahler and complex structures of the six-dimensional metric
 \be
 \d s^2_Y=e^{2D-\phi} \d s^2_6= (\d s^2_X+e^{-\phi} |h(z)|^2 \d z \d \bar z) \label{dsy}
 \ee
  Finally let us consider the remaining equation (\ref{susycond3}) and specialize to the backgrounds we are interested in. Then $Y=\mathbb{C}\times X$, where $X$ is a K3-space. We use complex coordinates $u^\alpha=(u^1,u^2,z)$  on $Y$
and write the differentials as
\bea
\label{ddc}
\d&=&\partial +\bar \partial    \qquad ~~~~~~~~~~~~~~  \partial=\d u^\alpha \wedge\partial_\alpha  \nn\\
\d^c &=& -\ii(\partial -\bar \partial ) \qquad ~~~~~~~~ \bar\partial=\d \bar u^\alpha\wedge \partial_{\alpha} 
\eea
The six-dimensional Hodge dual can be computed with the help of the formulas 
\bea
*_6\, \d f&=& \ft{1}{2}  \, \ J_{\rm st} \wedge J_{\rm st}  \wedge \d^c f \quad
*_6 \, (\ft12 \,J_{\rm st} \wedge J_{\rm st} \wedge \d f) =\,  \d^c f  \quad
*_6  \, \chi_a \wedge \d f  =  -  \chi_a \wedge \d^c f \nn
 \eea
valid for any function $f$ and any self-dual two-form $\chi_a$ on $X$.  
Two form will be always expanded in the basis of self-dual two forms $\chi_a= [\calc_a]$ associated 
 to a set of exceptional cycles $\calc_a$  at a singularity of K3. 

  Using this equations (\ref{susycond3})  can be written as  
 \bea
 F_1 &=& \ft12 e^{-4D} *_6     \d( e^{\phi}    J\wedge J) = -\d^c e^{-\phi}   \nn\\ 
 F_3 &=& e^{-\phi} *_6  \d B =-e^{-\phi} \d^c B   \nn\\
F_5 &=& -e^{-4D} *_6 \d (e^{4D-\phi} )  =\ft12 J\wedge J\wedge \d^c e^{\phi-4D}  \label{eqsa0}
  \eea
  Writing $F=\d (e^{B}\wedge C)$ and collecting $\partial$, $\bar\partial $-components the three equations can
  be written in the compact form
 \be
{\rm Re} ( \bar\partial {\cal T} )=0 \label{dtau}
 \ee
with\footnote{The polyform defined in (\ref{bardt}) is a specialized version of the polyform $\calt$ used in \cite{eff1,eff2} to discuss general warped flux compactifications.}
 \be
 {\cal T}  =e^B \wedge  (C+ \ii \,e^{-\phi}\, {\rm Re }\Psi )   \label{bardt}
  \ee 
  In components
  \bea
   {\cal T}_0 &=& \tau\equiv C_0+\ii \, e^{-\phi}  \nn\\
   {\cal T}_2  &=& C_2 + \tau B \equiv\beta^a\,\chi_a\nn\\
    {\cal T}_4 &=& C_4-  \ft{\ii}{2} e^{-4A}J\wedge J   
   +C_2\wedge B +  \ft12   \tau  B\wedge B  \label{t024fin}
    \eea  
     after setting  $e^{2D}=e^{2A+\frac{\phi}{2}}$.
         One can then easily see that the first two equations in (\ref{dtau})  can be solved by taking 
  $\tau$ and $\beta^a$ some holomorphic functions on $\mathbb{C}$: $\tau=\tau(z)$ and $\beta^a=\beta^a(z)$.
The holomorphicity of $\beta^a$ ensures the famous ISD condition on the three form field
   strength $*_YG_3=\ii G_3$, with $G_3= F_3+\ii e^{-\phi} H$,  where $*_Y$ is the Hodge operator associated  with the metric (\ref{dsy}). 
   
    The last equation in (\ref{dtau}) requires more care. Indeed, (\ref{dtau}) necessarily implies
     that $ \del\delbar\Im\calt_4=    \del\delbar\Re\calt_4=0$, which provides the following equations for the warping
   and the four-form field
     \bea\label{warpingf4}
 &&    \partial \bar \partial C_4= -  \partial \bar \partial ( C_2 \wedge B+\ft12 C_0 \, B\wedge B ) \nn\\
  && \del\delbar  (e^{-4A}) \wedge J\wedge J= \del\delbar  (  e^{-\phi} B \wedge B )    
  \eea
        
We write the ten-dimensional Einstein metric as
  \be
  \d s^2_{\rm E} = e^{-\frac{\phi}{2}} \d s^2_{\rm st}=e^{2A} \d x^\mu \d x^\mu+e^{-2A} \d s^2_Y 
  \ee 
which is somewhat more natural in this context and is then used in the main text. 


  Finally we remark that the structure of the solutions here is preserved under any SU(2)$_R$ transformation which rotates 
the three two-forms $(\Re \omega,\Im \omega,j)$. This   implies that the four-dimensional supersymmetry is enhanced to $\caln=2$.

\section{BPS solutions of the six-dimensional supergravity}
\label{app:trunc}

In this paper we construct  supersymmetric vacua of ${\cal N}=(2,0)$ six-dimensional supergravity in which a subset of fields vary over a complex plane with non-trivial holonomies under  a subgroup of the U-duality group.  In this Appendix we discuss the details of the moduli space and the truncations we use.  
  
 \subsection{Moduli space of type IIB supergravity on K3}
  \label{app:trunc1}
Let us first summarize some  properties of type IIB compactifications on K3 surfaces, referring to \cite{aspin} for more details. 
 The complete set of 105 moduli describing a compactification of type IIB supergravity on a K3 surface spans the orbifolded coset
 moduli space 
  \be
 \calm_{\text{IIB on K3}}={\rm O}(\Gamma_{5,21})\backslash {\rm O}(5,21;\mathbb{R})/ ({\rm O}(5;\mathbb{R}) \times {\rm O}(21;\mathbb{R})
  \label{modulispace}
  \ee
  A point in this space can be thought of as a choice of a time-like five-plane $\Pi\subset  \mathbb{R}_{5,21}$ with 
  ${\rm O}(5;\mathbb{R})$ and  ${\rm O}(21;\mathbb{R})$ acting as rotations along and perpendicular to $\Pi$ respectively. 
On the other hand ${\rm O}(\Gamma_{5,21})\simeq{\rm O}(5,21;\mathbb{Z})$ is the U-duality group which acts by 
rotations in $\mathbb{R}_{5,21}$ preserving  an even self-dual lattice
$\Gamma_{5,21}$ with a non-degenerate pairing $\cali$  of signature $(5,21)$. One can choose a basis of elements 
\be\label{latbasis}
\chi_\Sigma=\{\zeta^+_i,\zeta^-_{r},\chi_A\}\qquad~~ i,r=1,2\qquad~~ A=1,\ldots,22
\ee
in which the pairing $\cali$ takes the form
\be\label{genmetric}
{\cal I}_{\Sigma\Lambda}=\left(\begin{array}{ccc}
0 & \bbone_2 & 0\\
\bbone_2 & 0 & 0\\
0 & 0 & \cali_{AB}
\end{array}\right)
\ee
The elements $\chi_A$ span the even self-dual lattice 
 $\Gamma_{3,19}\simeq H^2(X;\mathbb{Z})$ of integer closed two-forms, with natural pairing given by $\cali_{AB}=\int_{X}\chi_A\wedge \chi_B$\footnote{We use an orientation convention  in which $\cali_{AB}$ has signature $(3,19)$, differently from the one used  for instance in \cite{aspin}.}.
  In turn, one may choose a basis  in which $\cali_{AB}=H^3\oplus \hat \cali$, where $H\equiv {\tiny \left(\begin{array}{cc} 0 & 1\\ 1 & 0\end{array}\right)}$  and $\hat\cali$ is the (positive definite) Cartan matrix of $E_8\times E_8$.

The moduli space (\ref{modulispace}) can be parametrized in terms of a vielbein $U_{\ul\Sigma}=U_{\ul \Sigma}{}^{\Lambda}\chi_\Lambda$ for $\mathbb{R}_{5,21}$,   where  `flat' indices are denoted by $\ul\Sigma,\ul\Lambda,\ldots$ and `curved' indices by $\Sigma,\Lambda,\ldots$. The $26\times 26$ matrix $U_{\ul \Sigma}{}^{\Lambda}$ satisfies
\be
U\, \cali \, U^T=\eta
\ee
where $\eta=  (-\bbone_5,\bbone_{21})$. In particular, the first five rows of the matrix $U$ span the five-plane $\Pi\subset  \mathbb{R}_{5,21}$ which contains the physical information. 
  The vielbein $U$ is defined up to `gauge' rotations $U\to O\, U$ with $O\in {\rm O}(5;\mathbb{R})\times {\rm O}(21;\mathbb{R})$.
 In addition,  the vielbeins $U$ and   $ U\, {\cal O} $ have to be identified for $ {\cal O}$ an element of the U-duality group
   ${\cal O}\in {\rm O}(5,21;\mathbb{Z})$, i.e.  $\calo \, \cali \, \calo^T=\cali$. 

 One can parametrize the gauge invariant degrees of freedom by using the matrix
  \be
  M\equiv U^T \, U
  \ee
    By construction $M$ is gauge invariant, symmetric $M^T=M$ and  satisfies $M\cali M\cali=\bbone$.  
   Furthermore, it transforms as $M\rightarrow \calo^T M\calo$ under $\calo\in {\rm O}(5,21;\mathbb{Z})$. 
   The matrix $M$ allows to define the sigma model which characterizes the six-dimensional effective action
\be\label{sigmamodel}
-\frac{M^4_{\rm P}}{8}\int \d^6x\sqrt{-g}\, {\rm tr}  \left( \cali  \, \nabla^N M\,\cali \, \nabla_N  M\right)  
\ee

%
%
  
In order to describe an explicit parametrization of the vielbein $U$, it is useful to introduce another $26\times 26$ matrix $V$ related to $U$ by
 \be
U=A V
\ee   
    with
  \be
 \label{diagmatrix}
A=\frac{1}{\sqrt{2}}\left(\begin{array}{ccc}
-\bbone_5 & \bbone_5 & 0  \\
 \bbone_5 & \bbone_5 & 0 \\
0 & 0 & \sqrt{2} {\cal E}
\end{array}\right)
\ee
and ${\cal E}$ a vielbein for the $E_8\times E_8$ pairing  $\hat\cali={\cal E}^T {\cal E}$. Then  an explicit  gauge-fixed parametrization of matrix $V$ is given by
\bea
\label{vv}
V &=&\left(\begin{array}{ccc}
E & -EC & -EY^T \\
0 & (E^{-1})^T & 0 \\
0 &  \hat\cali Y & \bbone
\end{array}\right) 
\eea
with
\be
  C= B+\ft12 Y^T\hat\cali \, Y   \qquad B=-B^T \qquad G=(E^T \, E)^{-1}
\ee
Here $G,B,C,E$ are $5\times 5$ matrices, Y is a $16\times 5$ block, while $\hat\cali$ is the $16\times 16$ matrix providing the $E_8\times E_8$ pairing. A similar gauge-fixed form of the  vielbein of the moduli coset space appears for example in \cite{schw92}, in the context of toroidal compactifications of the heterotic string on $T^5$ with $G$, $B$ and $Y$ corresponding to the metric, $B$-field
and Wilson lines respectively. 


\subsection{IIB identification of the moduli}
\label{app:10Dmoduli}

In order to clarify the ten-dimensional interpretation of  the description of the moduli space given above, it is convenient
to denote the first five (time-like) elements of the vielbein $U_{\ul\Sigma}$ by $(U_{\ul i},U_{\ul \alpha})$, $\ul i=1,2$, $\ul\alpha=1,2,3$.
Then, by using the lattice basis (\ref{latbasis}) one can choose $U_{\ul\alpha}$ to be of the form 
$U_{\ul\alpha}=U_{\ul\alpha}{}^i\zeta^+_i+U_{\ul\alpha}{}^A\chi_A$. 
In this gauge, the metric moduli, up to the overall volume of K3, are encoded in the three elements 
\be
J_{\ul\alpha}\equiv U_{\ul\alpha}{}^A\chi_A\in H^2(X;\mathbb{R})
\ee which generate a three-dimensional time-like plane  $\Sigma\subset H^2(X;\mathbb{R})$.  
More explicitly, the  elements   $J_{\ul\alpha}$  can be identified, up to the overall rescaling, 
with the triplet of real anti-self-dual harmonic two-forms 
 \be
 J_{\ul \alpha}=\ft{1}
 {  \sqrt{2\text{vol}_X}  } (\Re\omega,\Im\omega,j)
 \ee
  which define the hyperk\"ahler structure of the K3 surface, cf.~footnote \ref{foot:hyper}.
Hence the moduli space of  metrics (up to the overall volume) in K3 is spanned by the $57$-dimensional Grassmanian  submanifold of  $\calm_{\text{K3}}$ given by
 \be
 \calm_{\text{K3-metrics}}={\rm O}(\Gamma_{3,19})\backslash {\rm O}(3,19;\mathbb{R})/ ({\rm O}(3;\mathbb{R}) \times {\rm O}(19;\mathbb{R})
 \ee
 where ${\rm O}(\Gamma_{3,19})$ is the geometrical duality subgroup, which acts on $H^2(X;\mathbb{R})$ while preserving the metric $\cali_{AB}$ of the lattice $H^2(X;\mathbb{Z})$. 
 
 On the other hand, we can write $U_{\ul i}=U_{\ul i}{}^j(\zeta^+_j+\calb_j{}^{A}\chi_A)+U_{\ul i}{}^r\zeta^-_r$. Then the $2\times 22$ matrix  $\calb_j{}^{A} $ 
 parametrize the $44$ components  of the NSNS and RR two-forms  $B$ and $C_2$ respectively, while the matrices  $U_{\ul i}{}^j$ and $U_{\ul i}{}^r$ encode the information on  $\phi$, $C_0$, $C_4$ and the warp factor. Below we will provide an explicit parametrization of these fields 
 for the cases of interest for this paper.


\subsection{Truncation of the moduli space}

Let us first consider the simplest truncation in which we set $B=C_2=0$. This corresponds to setting $U_{\ul i}{}^A=0$ 
which reduces the matrix $U_{\ul \Sigma}{}^\Lambda$ into a block diagonal form with blocks of dimensions $4$ and $22$.
We can then truncate the dynamical fields by keeping fixed the metric components $U_{\ul \alpha}=U_{\ul \alpha}{}^A\chi_A$, while allowing a dynamical $U_{\ul i}=U_{\ul i}{}^I\zeta_I$, where we have introduced $\zeta_I=(\zeta^+_i,\zeta^-_r)$, $I=1,\ldots, 4$. The two vectors $U_{\ul i}$ span a two-dimensional time-like subplane of $\mathbb{R}^{2,2}$ and the restricted moduli space is then given by 
\be\label{trunc1}
\calm={\rm O}(\Gamma_{2,2})\backslash {\rm O}(2,2;\mathbb{R})/{\rm O}(2;\mathbb{R})\times {\rm O}(2;\mathbb{R})
\ee
 We notice that this space is invariant under rotations in the orthogonal subgroup ${\rm O}(3,19;\mathbb{R})$ and therefore
 defines a consistent truncation of the scalar moduli space.
 
We would like now to extend the truncation (\ref{trunc1}) by including non trivial NSNS and RR fluxes. 
More precisely, we would like to allow for more general dynamical   $U_{\ul i}$'s, while still keeping the $U_{\ul\alpha}$'s fixed. This can be achieved as follows. We first take a set of $n$ space-like integer closed forms $\chi_a\in H^2(X;\mathbb{Z})$, $a=1,...n$, with a non-degenerate positive definite pairing $\Delta_{ab}=\int \chi_a\wedge \chi_b$. Let us also assume that there are other $22-n$ integer forms $\tilde\chi_{\tilde a}$ which are orthogonal to the $\chi_a$'s, i.e.\ $\int_X\chi_a\wedge \tilde\chi_{\tilde b}=0$. Together $(\chi_a,\tilde\chi_{\tilde a})$ span $H^2(X;\mathbb{R})$ but in general generate a sublattice of $H^2(X;\mathbb{Z})$. 

Then, the truncation is specified by restricting $U_{\ul\alpha}=U_{\ul\alpha}{}^{\tilde a}\tilde \chi_{\tilde a}$ and $U_{\ul i}=U_{\ul i}{}^{j}\zeta^+_j+U_{\ul i}{}^r\zeta^-_r+U_{\ul i}{}^{a}\chi_a$ and allowing only $U_{\ul i}$ to be dynamical, while  $U_{\ul\alpha}$ is kept fixed. 
In other words we focus on the scalar fields which specify a two-dimensional plane $\Pi_2$ spanned by the 
 two vectors $U_{\ul i}$ in  the space $\mathbb{R}^{2,2+n}\simeq \Gamma_{2,2+n}\otimes \mathbb{R}$, where $\Gamma_{2,2+n}=\Gamma_{2,2}\oplus \Gamma_n$ is the lattice spanned by $(\zeta^+_i,\zeta^-_r,\chi_a)$.  Our truncation is then given by a block-diagonal complete vielbein $U_{\ul\Sigma}{}^\Lambda$, with dynamical blocks of dimensions $(4+n)$ and 
 a constant block of dimension $(22-n)$. 
 
Clearly, we can use the description of the dynamical moduli in terms of the coset   
\be\label{redmoduli}
\calm={\rm O}(\Gamma_{2,2+n})\backslash {\rm O}(2,2+n;\mathbb{R})/{\rm O}(2;\mathbb{R})\times {\rm O}(2+n;\mathbb{R})
\ee
where ${\rm O}(\Gamma_{2,2+n})$ is the subgroup of ${\rm O}(\Gamma_{5,21})$ which acts only on the basis $(\zeta^+_i,\zeta^-_r,\chi_a)$
and leaves $\tilde \chi_{\tilde a}$ untouched.

  Notice that the above ansatz requires that $\int J_{\ul\alpha}\wedge \chi_a=0$. According to the discussion provided in section \ref{app:10Dmoduli}, this means that   
\be\label{ortcond}
\int_{\calc_a}j=\int_{\calc_a}\Re\omega=\int_{\calc_a}\Im\omega=0
\ee 
where $\calc_a$ are the cycles which are dual to the integer closed forms $\chi_a\in H^2(X;\mathbb{Z})$.
The condition (\ref{ortcond}) implies that the cycles $\calc_a$  dual to $\chi_a$  must have vanishing volume. 
Indeed, it is known that a K3 surface develops an orbifold singularity if an only if the three-plane  $\Sigma$ is orthogonal to some points of the lattice $H^2(X;\mathbb{Z})$, see for instance \cite{aspin}. The cycles    $\calc_a$ are just the exceptional cycles associated with the orbifold singularity.\footnote{Notice however that $\Pi$ itself is generically not orthogonal to any element of the lattice and then the theory is not singular \cite{aspin}.}  As a concrete example, one can consider $X$ to be $T^4/\mathbb{Z}_2$ blown-up at $16-n$ points.  We take $\chi_a$ to be Poincar\'e dual to the $n$ unresolved  exceptional cycles, while the remaining $22-n$ two-forms $\tilde\chi_{\tilde a}$ can be taken to be  the Poincar\'e duals to the remaining $16-n$ (blown-up) exceptional cycles and of the $6$ toroidal cycles inherited from the underlying $T^4$.\footnote{See for instance \cite{Kumar:2009zc,Schulz:2012wu} for readable discussions on the structure of $H^2(X;\mathbb{Z})$ and some of its sublattices in the case of Kummer K3 spaces. }

\subsection{Holomorphic parametrization}  
\label{app:gaugefix}

Let us now give an explict parametrization of the $(4+n)\times (4+n)$ vielbein $U$  describing the reduced moduli space coset (\ref{redmoduli}). We use the same strategy used for the general case in section \ref{app:trunc1}. Namely, we introduce a matrix $V$ related to $U$ by
 \be
U=A V
\ee   
    with
  \be
 \label{diagmatrix2}
A=\frac{1}{\sqrt{2}}\left(\begin{array}{ccc}
-\bbone_2 & \bbone_2 & 0  \\
 \bbone_2 & \bbone_2 & 0 \\
0 & 0 & \sqrt{2} {\cal E}
\end{array}\right)
\ee
where ${\cal E}$ is now a vielbein for the positive definite $\Delta_{ab}$:  $\Delta={\cal E}^T {\cal E}$. 
We can then use a gauge-fixed parametrization on $V$ analogous to (\ref{vv}), with $\hat\cali$ substituted by $\Delta$, with $E,B,C$ now $2\times 2$ matrices and $Y$ a $n\times 2$ block.  
We can then introduce the complex fields $\tau,\sigma,\beta^a$ defined by
 \be\label{vhol}
\begin{aligned}
E    &=\frac{1}{\sqrt{\calv}}\left(\begin{array}{cc}
\Im\tau & 0 \\
 -\Re\tau & 1
\end{array}\right)\\
\beta^a&=-Y_2{}^a+\tau Y_1{}^b\\
\Re\sigma&=B_{12}+\frac1{2\Im\tau}\Re\beta\cdot\Im\beta\\
\end{aligned}
\ee
where 
\be
\calv= \Im\tau\Im\sigma-\frac12\Im\beta\cdot\Im\beta\\
\ee
and we have used a notation in which, for instance, $\Im\beta\cdot\Im\beta=\Delta_{ab}\Im\beta^a\Im\beta^b$.  

One can now compute the coset `metric' (\ref{genmetric}) and using it in the sigma model (\ref{sigmamodel})
one finds
\be
-\ft14 
 \, {\rm tr}  \left( \cali  \, \nabla^N M\,\cali \, \nabla_N  M\right)  
 =-2 K_{I\bar J}(\varphi)\nabla_M\varphi^I\nabla^M\bar\varphi^{\bar J} 
 \ee
  with $\varphi^I=(\tau,\sigma,\beta^a)$, $K_{I\bar J}={\partial^2 \over \partial \varphi^I \varphi^{\bar J}} K$ and $K$ the 
  K\"ahler potential 
\be\label{kah2}
K=-\log\calv
\ee
Namely,  the effective action for the truncated scalar sector coupled to gravity  takes the form (\ref{effact}).  
 
\subsection{Supersymmetry analysis}
\label{app:susy}

In this section we explicitly study the supersymmetry properties of the vacua considered in this paper, from the effective six-dimensional perspective. 

In general, a supersymmetric bosonic vacuum must satisfy the Killing spinor equations, obtained by imposing the vanishing of the supersymmetry variations of the fermionic fields. In the case of a compactification of  IIB supergravity on K3,
the effective six-dimensional theory is given by an $\caln=(2,0)$ supergravity coupled to $21$ self-dual tensor multiplets. The general structure of $\caln=(2,0)$ supergravities has been determined in \cite{romans} and another useful reference, whose conventions we follow, is provided by \cite{gutperle}. The fermionic fields are given by the gravitino $\psi_M$ and the 21 tensor multiplet fermions $\rho_{\ul{\hat r}}$, ${\ul{ \hat r}}=1,\ldots,21$, which carry the (four-dimensional) spin representation of the ${\rm SO}(5;\mathbb{R})$ R-symmetry group. Furthermore, the index ${\ul{\hat r}}$ of $\rho_{\ul{\hat  r}}$ transforms in the fundamental representation of ${\rm SO}(21;\mathbb{R})$. The fermions  $\psi_M$ and  $\rho_{\ul{\hat r}}$ have opposite chirality
\be
\Gamma_7\psi=-\psi\qquad~~~~\Gamma_7\rho_{\ul{ \hat r}}=\rho_{\ul{ \hat r}}
\ee
with $\Gamma_7=\Gamma^{012345}$, and satisfy the symplectic-Majorana conditions 
$\psi_M=\calc\hat\calc(\psi_M)^*$ and $\rho_{\ul r}=\calc\hat\calc(\rho_{\ul{\hat r}})^*$,
where $\calc$ and $\hat\calc$ are complex conjugation matrices for the spin representations of the space-times SO$(1,5;\mathbb{R})$ and the R-symmetry SO$(5;\mathbb{R})$, respectively. More explicitly, denoting the  R-symmetry SO$(5;\mathbb{R})$ gamma matrices by $\gamma^{\ul r}$, 
$\calc$ and $\hat\calc$ are defined by
\be
\calc\Gamma_M\calc^{-1}=(\Gamma_M)^* \quad~~~~~~ \hat\calc\gamma^{\ul r}\hat\calc^{-1}=(\gamma^{\ul r})^*
\ee

We are interested in six-dimensional backgrounds with non-trivial scalars but vanishing three-form fluxes. Hence, the relevant supersymmetry transformations reduce to 
\be\label{gensusy}
\begin{aligned}
\delta\psi_M&=\big[\nabla_M\epsilon-\frac{1}{4}(Q_M)_{\ul{rs}}\,\gamma^{\ul{rs}}\big]\epsilon\\
\delta\rho_{\ul{\hat r}}&=\frac1{\sqrt{2}}\Gamma^M (P_M)_{\ul {\hat r s}}\,\gamma^{\ul{s}}\epsilon
\end{aligned}
\ee 
where $\epsilon$ has the same spinorial property as $\psi_M$:
\be\label{epsiloncond}
\Gamma_7\epsilon=-\epsilon\, ,\quad~~~
\epsilon=\calc\hat\calc\epsilon^*
\ee
 In (\ref{gensusy}) the matrices $Q_M$ and $P_M$  are given  by the formula
\be
\partial_M U\, U^{-1}=\left(\begin{array}{cc} Q_M & \sqrt{2}P_M \\
\sqrt{2}P^T_M & S_M\end{array}\right)
\ee 
where $\{U_{\ul\Sigma}\}=\{U_{\ul r},U_{\ul{\hat r}}\}$ is  the coset vielbein introduced in Appendix \ref{app:trunc1}. Notice that the indices ${\ul r}$ and $\ul{\hat r}$, being flat, can be raised and lowered with no problems. In this sense, $Q^T_M=-Q_M$.

We can now evaluate $Q_M$ and $P_M$ for scalars belonging to the truncated moduli space (\ref{redmoduli}), 
 by using the gauge-fixed vielbein provided in Appendix \ref{app:gaugefix}, and plugging them in the supersymmetry transformations (\ref{gensusy}).

Let us  consider more in detail the R-symmetry connection $Q_M$, which appears in the gravitino supersymmetry condition. Clearly, the only non vanishing components of $Q_M$ are $Q_{M\ul{ij}}=\calq_{M}\epsilon_{\ul{ij}}$, which can be read from
 \be
\calq_{M}\,\d x^M=\Im\left( \frac{\del K}{\del\varphi^I}\,\partial_M\varphi^I\right)
 \label{QM}
 \ee
 where $K$ is defined in (\ref{kah2}). We notice that  $\calq_{M} $ 
  is the pull back of the of the U(1) connection on $\calm$ associated with
   the holomorphic line bundle whose sections transform as modular forms of weight one.

    The gravitino supersymmetry transformations can then be written as
\be
\delta\psi_M=\big[\nabla_M-\frac{1}{2} \calq_{M} \gamma^{\ul{12}}  \big]\epsilon
\ee
 
\subsubsection{Supersymmetric vacua}

Let us now focus on our vacua, which are characterized by the six-dimensional metric
\be
\d s^2=\d x^\mu\d x_\mu+M_P^{-4}\,  \calv |h(z)|^2\d z\d \bar z
\ee
and complex scalars $\varphi^I=(\tau,\sigma,\beta^a)$ depending holomorphically just on $z$: $\delbar\varphi^I(z)=0$.

In order to prove that these vacua preserve four-dimensional $\caln=2$ supersymmetry\footnote{See for instance \cite{bergsh,howe} for analogous discussions on supersymmetric codimension-two configurations.}, we can use the following representation of the SO$(1,5;\mathbb{R})$ gamma matrices:
\be
\Gamma^{\mu}=\hat\gamma^\mu\otimes \bbone\, ,\quad~~~ \Gamma^{5}=\hat\gamma_5\otimes \sigma_1\, ,\quad~~~ \Gamma^{6}=\hat\gamma_5\otimes \sigma_2
\ee
where $\hat\gamma^\mu$ are four-dimensional gamma-matrices, which we choose to be real, and $\hat\gamma_5\equiv -\ii\hat\gamma^{0123}$ is the associated chiral operator. In this representation $\Gamma_7=-\hat\gamma_5\otimes\sigma_3$ and we can take $\calc=\bbone\otimes\sigma_2$.
Analogously, we can take the following explicit (four-dimensional) representation of the  SO$(5;\mathbb{R})_{\rm R}$ R-symmetry  (in this section we introduce the suffix $_R$ for clarity) gamma matrices $\gamma^{\ul r}=(\gamma^{\ul i},\gamma^{\ul\alpha})$, adapted to the decomposition ${\rm SO}(5;\mathbb{R})_{\rm R}\rightarrow {\rm SO}(2;\mathbb{R})_{\rm R}\times {\rm SO}(3;\mathbb{R})_{\rm R}$:
\be
\begin{aligned}
\gamma^{\ul i}&=(\hat\sigma_1\otimes \bbone,\hat\sigma_2\otimes \bbone)\\
\gamma^{\ul \alpha}&=(\hat\sigma_3\otimes \sigma_1,\hat\sigma_3\otimes \sigma_2,\hat\sigma_3\otimes \sigma_3)
\end{aligned}
\ee
where the $\hat\sigma_i$ are Pauli matrices acting on the spin representation of the the R-symmetry group,    
with associated charge conjugation matrix $\hat\calc=\hat\sigma_1\otimes\hat\sigma_2$.

Then, one can take the following spinorial ansatz for the six-dimensional spinor, which automatically satisfies (\ref{epsiloncond}):
\be\label{6dsusy}
\epsilon=\zeta\otimes\eta+(\sigma_1\zeta^*)\otimes (\sigma_2\hat\sigma_2\eta^*)
\ee
Here $\zeta$ is an arbitrary four-dimensional constant chiral spinor ($\hat\gamma_5\zeta=\zeta$) which transforms as a spin-doublet under the ${\rm SO}(3)_{\rm R}\simeq{\rm SU}(2)_{\rm R}$ R-symmetry sub-group. Hence, it has eight independent components, which correspond to the eight $\caln=2$ four-dimensional supercharges. On the other hand, in (\ref{6dsusy}) $\eta$ is a two-dimensional chiral spinor ($\sigma_3\eta=\eta$) which is chiral under $ {\rm SO}(2)_{\rm R}$ too: $\hat\sigma_3\eta=\eta$. 

Under these conditions, the gravitino supersymmetry condition reduces to an equation on the internal two-dimensional space:
\be\label{2dsusy}
(\nabla_m-\frac{\ii}{2} \calq_{m}  )\eta=0
\ee
where  $m$ runs over coordinates of the transversal complex plane. In (\ref{2dsusy}) $\nabla_m$ must be computed by 
using the two-dimensional metric $\calv |h(z)|^2\d z\d \bar z$. Hence, by taking into account (\ref{QM}), it is not difficult to see that in (\ref{2dsusy}) the $\cal V$-dependent terms cancel between the spin-connection and U(1) connections\footnote{Explicitly 
$\calq_z= -\frac{\ii}{2} \partial_z K $,  $\calq_{\bar z}=\calq_{z}^*$ and $\gamma^{\ul{12}} \eta=\ii \eta$. The non trivial components of the spin connection
 are $w_{z 1 2}=\ft{\ii}{2}\partial _{z} \log \left( \calv |h|^2 \right)$ and $w_{\bar{z} 12}=-\ft{\ii}{2}\partial _{\bar{z}} \log \left( \calv |h|^2 \right)$. } and
 one is left with a simple equation satisfied by
\be
\eta=\left(\frac{h(z)}{\bar h(\bar z)}\right)^{\frac14}\eta_0
\ee
with constant $\eta_0=\sigma_3\eta_0=\hat\sigma_3\eta_0$.

 The remaining supersymmetry conditions $\delta\rho_{\ul i}=\delta\rho_{\ul a}=0$ can be analyzed along the same lines. Namely,  one can compute  $P_M$ from the truncated vielbein, express the result in terms of the complex fields $(\tau,\sigma,\beta^a)$, use the above   explicit representations of the space-time and  R-symmetry gamma matrices, and evaluate $\delta\rho_{\ul i},\delta\rho_{\ul a}$ for the spinorial ansatz described around (\ref{6dsusy}). The result is that, indeed, $\delta\rho_{\ul i}=\delta\rho_{\ul a}=0$ once the complex fields $\tau,\sigma,\beta^a$ are chosen to depend  holomorphically on $z$.

 \section{Genus two curves, modular forms and theta functions}
 \label{sgtwo}
 
In this Appendix we summarize some useful facts about hyperelliptic curves of genus two. 
Although the discussion will be necessarily incomplete, we will try to be self-consistent in treating the results
useful for the present paper.  For more details the reader should consult, for instance, \cite{munford, farkas, bertola,Minahan:1995er, D'Hoker:2001qp}\footnote{We thank M. Bill\'o for notes  and explanations contributing to the presentation here.}. 

 \subsection{Genus two hyperelliptic curves}
 \label{AppCycles}
 
 A genus two surface $\Sigma$ can be described by a hyperelliptic curve in $\Cint^2$
 given by
\be
y^2=a_0\prod_{i=1}^6 (x-e_i)  \label{sucurve}
\ee
The curve can be interpreted in terms of a two-sheet covering of the 
$x$-complex plane with three cuts pairing the $e_i$'s, let us say along $[e_{2i-1},e_{2i}]$ with $i=1,2,3$. This description maps a point $p\in \Sigma$ to a point $x(p)\in  \mathbb{C}  $. 
\begin{figure}
    \begin{center}
    \def\svgwidth{7.5cm}
    \begingroup%
  \makeatletter%
  \providecommand\color[2][]{%
    \errmessage{(Inkscape) Color is used for the text in Inkscape, but the package 'color.sty' is not loaded}%
    \renewcommand\color[2][]{}%
  }%
  \providecommand\transparent[1]{%
    \errmessage{(Inkscape) Transparency is used (non-zero) for the text in Inkscape, but the package 'transparent.sty' is not loaded}%
    \renewcommand\transparent[1]{}%
  }%
  \providecommand\rotatebox[2]{#2}%
  \ifx\svgwidth\undefined%
    \setlength{\unitlength}{360.01045375bp}%
    \ifx\svgscale\undefined%
      \relax%
    \else%
      \setlength{\unitlength}{\unitlength * \real{\svgscale}}%
    \fi%
  \else%
    \setlength{\unitlength}{\svgwidth}%
  \fi%
  \global\let\svgwidth\undefined%
  \global\let\svgscale\undefined%
  \makeatother%
  \begin{picture}(1,0.71111536)%
    \put(0,0){\includegraphics[width=\unitlength]{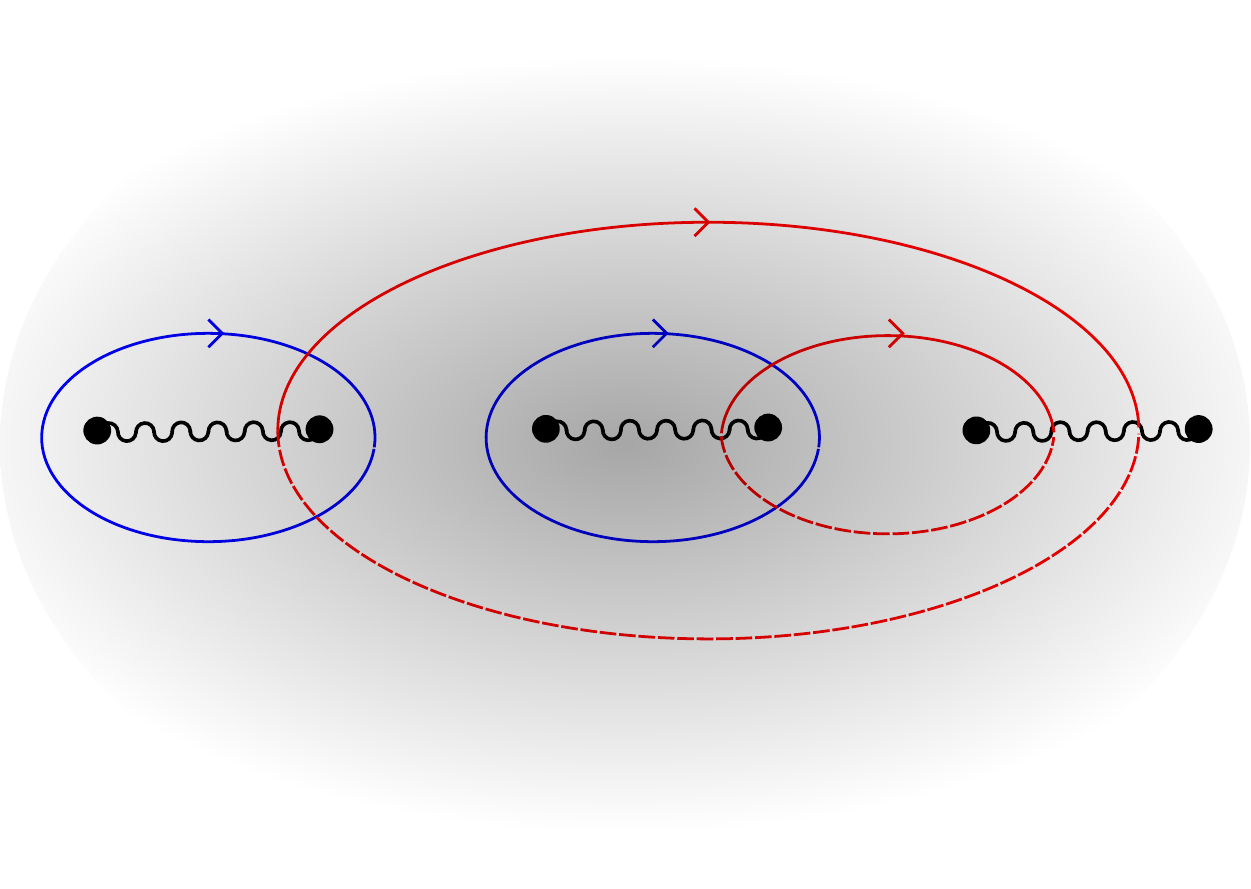}}%
    \put(0.05556548,0.32222531){\makebox(0,0)[lb]{\smash{$e_1$}}}%
    \put(0.23333809,0.32222531){\makebox(0,0)[lb]{\smash{$e_2$}}}%
    \put(0.41111071,0.32222531){\makebox(0,0)[lb]{\smash{$e_3$}}}%
    \put(0.59999411,0.32222531){\makebox(0,0)[lb]{\smash{$e_4$}}}%
    \put(0.76665594,0.32222531){\makebox(0,0)[lb]{\smash{$e_5$}}}%
    \put(0.92220698,0.32222531){\makebox(0,0)[lb]{\smash{$e_6$}}}%
    \put(0.07778705,0.44444399){\makebox(0,0)[lb]{\smash{$\gamma^1$}}}%
    \put(0.45555386,0.44444399){\makebox(0,0)[lb]{\smash{$\gamma^2$}}}%
    \put(0.48221975,0.54444108){\makebox(0,0)[lb]{\smash{${\tilde\gamma}_1$}}}%
    \put(0.66665884,0.46666556){\makebox(0,0)[lb]{\smash{${\tilde\gamma}_2$}}}%
  \end{picture}%
\endgroup%
    \end{center}
  \caption{A basis of cycles satisfying $\gamma^a\cdot\tilde\gamma_b=\delta^a_b$ for the genus two Riemann surface. The points $e_i$ represent the six roots
of the curve in the $x$ plane  (Fig. in \cite{lerdaetnoi}).}
\label{cycles}
\end{figure}
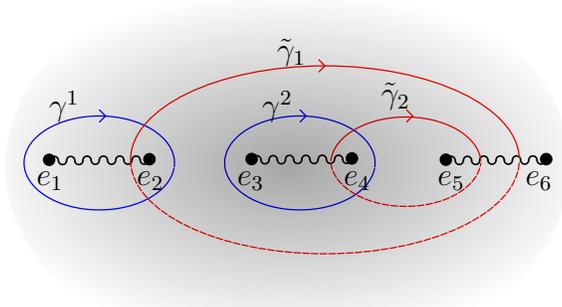
The first homology class of $\Sigma$ has dimension $b_1(\Sigma)=4$ and one can choose a symplectic basis of one-cycles $\{\gamma^a,\tilde\gamma_b\}_{a,b=1,2}$, which have intersection numbers $\gamma^a\cdot\tilde\gamma_b=\delta^a_b$. In the double-sheet description provided by (\ref{sucurve}), one can make the following choice. 
The cycle $\gamma^a$ encircles clockwise the cuts $[e_{2a-1},e_{2a}]$ in one sheet, while  $\tilde\gamma_a$ goes along one sheet from $[e_{2a-1},e_{2a}]$ towards $[e_{5},e_{6}]$ and comes back  along the second sheet, see figure \ref{cycles}. It is easy to check that indeed $\gamma^a\cdot \tilde\gamma_b=\delta^a_{b}$ for this choice. 
 
The curve $\Sigma$ is charactered by its {\em period matrix} 
\be
{\bf \Omega}=\left(\begin{array}{cc} \tau & \beta \\ 
\beta & \sigma \end{array}\right)
\ee  
 defined as follows. Take a basis of holomorphic one-forms $\lambda_a$, $a=1,2$, on $\Sigma$. Then, 
  \be
  \label{MN}
 {\bf \Omega}_{ab}=(N M^{-1})_{ab}\, \qquad\qquad \text{with}  \qquad 
 M^a{}_{b}   =\frac{1}{2\pi\ii}\oint_{\gamma^a} \lambda_b 
 \  \quad~~
 N_{ab}  =\frac{1}{2\pi\ii}\oint_{\tilde \gamma_a} 
     \lambda_b    \ee
Alternatively, one can introduce the normalized holomorphic differentials $\omega_a=\lambda_b(M^{-1})^b{}_a$, $\oint_{\gamma^a}\omega_b=\delta^a_b$ and define ${\bf \Omega}_{ab}=\oint_{\tilde\gamma_a}\omega_b$.   Notice that by construction ${\bf \Omega}_{ab}$ is symmetric and has positive definite imaginary part.

In our setting, we can choose the basis $\lambda_a=\frac{x^{a-1} {\rm d}x}{y }$ and, by using the one-cycles described above, the
period integrals reduce to line integrals with the identifications 
    \be
\oint_{\gamma^a} \lambda_b= 2\int_{e_{2a-1}}^{e_{2a}} \lambda_b \  , \qquad~  \oint_{\tilde{\gamma}_a} \lambda_b=2\sum_{p=a}^2 
\int_{e_{2p}}^{e_{2p+1}} \lambda_b   \label{periods}
\ee

The definition of $\Omega$ depends on the choice of the symplectic basis $\{\gamma^a,\tilde\gamma_b\}$ and, clearly, any re-shuffling of such one-cycles should produce an equivalent period matrix. The most general redefinition of symplectic basis corresponds to an element of ${\rm Sp}(4,\mathbb{Z})$, the {\em modular group}, which is parametrized as in (\ref{spmatrix}). More explicitly, it is defined by four $2\times 2$ matrices  $(A_a{}^b,B_{ab}, C^{ab}, D^a{}_b)$, which take values in $\mathbb{Z}$ and satisfy the constraints 
  \be
  A^T C=C^T A \qquad  B^T D=D^T B \qquad A^T D-C^T B=D^T A-B^T C=\bbone
  \ee
The modular group acts on the period matrix ${\bf\Omega}_{ab}$  by
\be
{\bf\Omega}\rightarrow (A{\bf\Omega}+B)(C{\bf\Omega}+D)^{-1}
\ee 
As a simple example, consider the basis change  $\gamma^a\rightarrow \gamma^a$, $\tilde\gamma_a\rightarrow \tilde\gamma_a+n_{ab}\gamma^b$, with $n_{ab}=n_{ba}\in\mathbb{Z}$. This generates the shifts ${\bf\Omega}_{ab}\rightarrow {\bf \Omega}_{ab}+n_{ab}$.


\subsection{Modular forms}

In our discussions, an important role is played by the {\em modular forms}, defined as follows.    A modular form $f({\bf \Omega})$ of weight $k$ is a function transforming as 
\be
f(\tilde{\bf \Omega})={\rm det} (C {\bf \Omega}+D)^{k} f({\bf\Omega})
\ee
where $\tilde{\bf\Omega}=(A{\bf\Omega}+B)(C{\bf\Omega}+D)^{-1}$.
The ring of modular forms is generated by the Eisenstein series defined as
\be
\psi_k({\bf \Omega}) = \sum_{C,D} {\rm det} (C {\bf\Omega}+D)^{-k}
\ee
where the sum is taken over the set of bottom halves $(C,D)$ of elements 
of the ${\rm Sp}(4,\Zint)$ group.

Any intrinsic property of the curve should be invariant under modular transformations.
Indeed, one can characterize the genus two surface by the so called {\em absolute Igusa invariants}  
\be
j_1={I_2^5\over I_{10} } \qquad  j_2={I_2^3 I_{4} \over I_{10}} \qquad j_3={I_4^2 I_{2}\over I_{10}}
\ee
given in terms of the  polynomials, known as the {\em homogeneous Igusa-Clebsch invariants}  
\bea
\begin{aligned}
&I_2 =a_0^2 \sum_{15~{\rm perms}} e_{12}^2 e_{34}^2 e_{56}^2 \\
&I_{4} =a_0^4 \sum_{10~{\rm perms}} e_{12}^2 e_{23}^2 e_{31}^2 e_{45}^2 e_{56}^2 e_{64}^2\\
&I_{6} = a_0^6\sum_{60~{\rm perms}} e_{12}^2 e_{23}^2 e_{31}^2 e_{45}^2 e_{56}^2 e_{64}^2
e_{14}^2 e_{25}^2 e_{36}^2   \\
 &I_{10} =a_0^{10} \prod_{1\leq i<j \leq 6} e_{ij}^2\\
 &I_{15} = \prod_{15~{\rm perms}} {\rm det} 
\left(
\begin{array}{ccc}
 1 & e_1+e_2  & e_1 e_2   \\
 1 & e_3+e_4  & e_3 e_4   \\ 
 1 & e_5+e_6  & e_5 e_6   \\
\end{array}
\right)
\end{aligned}
   \label{iis}
\eea
generating the ring of projective invariants. $I_{10}$ is the discriminant of the curve and for $I_{10}\neq 0$ the Riemann surface is smooth\footnote{The  discriminant of a hyperelliptic curve $y^2=a_0 \prod_{i=1}^{2n} (x-e_i)$ of genus $g=n-1$ is defined 
as  $\Delta=a_0^{4n-2} \prod_{i<j} e_{ij}^2$ and it is invariant under the $SL(2,\Rint)$ transformations (\ref{sl2r}) .} . 
The sums in (\ref{iis}) run over the $15$ partitions 
into three groups of two elements, $10$ partitions  into two groups of three elements and 
$60=10\times 6 $ matching between two groups of three elements (10 choices for the partition
into two groups and six matching between the two chosen groups).\\
The $I_k$ polynomials defined above are invariant under a generic SL(2,$\Rint$) transformation in the $x$ plane
\be
\tilde x= \frac{a\, x +b}{c\,  x+d}~~~~~~~~~~~~~~~ \tilde y= \frac{ y}{(c\, x+d)^3}  \label{sl2r}
\ee
that maps the roots $e_i$ and $a_0$ according to
\be
 \tilde e_i = \frac{a\, e_i +b}{c\,  e_i+d}~~~~~~~~~~~ \tilde a_0=  a_0\prod_{i=1}^6(c\,  e_i+d)
\ee
 In terms of the invariants (\ref{iis})  one can write the basic
Siegel modular forms
\bea
\psi_4=\ft14 I_{4} \quad ~\psi_6=\ft18(I_{2} I_{4}-3 I_{6} )  \quad ~
\chi_{10}=-\ft{1}{2^{14}} I_{10} \quad ~\chi_{12}=
\ft{1}{2^{17} 3 } \, I_2\,  I_{10} \quad ~\chi_{35}=5^3\, I_{10}^2 I_{15}\nn\\
\label{siegelf}
\eea
with $\psi_{4}, \psi_{6}$ Eisenstein series of weight $4,6$ and $\chi_{10}, \chi_{12},\chi_{35}$ cusp forms 
of weight $10$, $12$ and $35$ respectively.  A cusp form $\chi_{k}$ is a modular form of modular weight $k$ which satisfies the condition \cite{geer}
\be
\lim_{ \sigma\to \ii \infty , \beta=0 } \chi_{k}({\bf \Omega})=0
\ee 
The modular forms defined in (\ref{siegelf}) generate the graded ring of classical Siegel modular forms of genus two.

 \subsection{The Abel map and Jacobi variety}

Introduce  the following vectors 
\be\label{C2basis}
v^1\equiv (1,0)\in\mathbb{C}^2\, ,\quad v^2\equiv(0,1)\in\mathbb{C}^2
\ee
and consider the map from $\Sigma$ to $\mathbb{C}^2$ defined by
\be\label{abel}
\phi(P)=\left(\int^{x_P}_{e_1}\omega_a\right) v^a
\ee
where we denote by  $x_P \in \Cint$ the projection of the point $P\in \Sigma$ on the $x$-plane, 
 the double-sheet description provided by (\ref{sucurve}).
If we shift $P$ along a general one-cycle $m_a\gamma^a+n^a\tilde\gamma_a$, then $\phi(p)\rightarrow \phi(p)+(m_a+{\bf \Omega}_{ab}n^b)v^a$. We then see that (\ref{abel}) defines a well defined map $\phi:\Sigma\rightarrow \mathbb{C}^2/\Lambda$, the {\em Abel map}, 
where $\Lambda$ is the $\mathbb{Z}^2\subset \mathbb{C}^2$ lattice generated by the vectors $v^a$ and $\tilde v_a\equiv \Omega_{ab}v^b$:
\be
\Lambda=\{m_av^a+n^b\tilde v_b| m_a,n^b\in\mathbb{Z}\}\, ,\qquad~~~\tilde v_a\equiv \Omega_{ab}v^b
\ee
The Abel map takes values into $\mathbb{C}^2/\Lambda$, which is the so-called {\em Jacobian variety}. 

The elements of the lattice $\Lambda$ are called `periods'. We introduce the following notation for the {\em half-periods}:
\be\label{halfper}
\left(\begin{array}{c} n \\ m\end{array}\right)\equiv \frac12(m_av^a+n^b\tilde v_b)
\ee
As elements of the Jacobian variety $\mathbb{C}^2/\Lambda$, the half-periods are $2\times 2$ matrices with entries 0 or 1.  
Then, in this notation, by using (\ref{periods}) one can express the value of the Abel map at the branch points $P_i\in \Sigma$ as follows 
\be\label{branchabel}
\begin{aligned}
&\phi(P_1)=\left(\begin{array}{cc} 0 & 0 \\ 0 & 0\end{array}\right)\quad\quad \phi(P_2)=\left(\begin{array}{cc} 0 & 0 \\ 1 & 0\end{array}\right)
\quad\quad \phi(P_3)= \left(\begin{array}{cc} 1 & 1 \\ 1 & 0\end{array}\right)\\
&\phi(P_4)= \left(\begin{array}{cc} 1 & 1 \\ 1 & 1\end{array}\right)\quad\quad \phi(P_5)=\left(\begin{array}{cc} 1 & 0 \\ 1 & 1\end{array}\right)
\quad\quad \phi(P_6)= \left(\begin{array}{cc} 1 & 0 \\ 0 & 0\end{array}\right)
\end{aligned}
\ee

\subsection{Theta functions}

 The  theta functions associated with  a genus two Riemann surface with period matrix ${\bf \Omega}$
 are defined as
 \be
 \theta [^a_b ](Z|{\bf\Omega})=\sum_{n\in \Zint^2} e^{ \pi \ii \left[ \left(n+{a\over 2} \right){\bf \Omega} \left(n+{a\over 2}\right)^T
 +2 \left(n+{a\over 2} \right)\left(Z+{b\over 2}\right)^T\right]}
 \ee
 Here $Z\equiv Z_a v^a\in \mathbb{C}^2$ [recall (\ref{C2basis}) for  the definition of $v^a$], $n\equiv (n^1,n^2)\in \Zint^2$, $a\equiv (a^1,a^2)\in \mathbb{Z}^2$ and $b\equiv (b_1,b_2)\in \mathbb{Z}^2$. The matrix $[^a_b]$ is called {\em half-characteristics}. Furthermore, we have used a notation in which, for instance, $n{\bf \Omega}n^T=n^a{\bf \Omega}_{ab}n^b$ and $nZ^T=n^aZ_a$. The half-characteristics $[^a_b]$ is called even/odd if $ab^T$ is even/odd respectively. 
 
The theta functions obey the following important properties:
 \begin{subequations}
 \begin{align}
  \theta[^{a}_{b} ](-Z|{\bf \Omega}) &= (-)^{ab^T} \,\theta[^{a}_{b} ](Z|{\bf \Omega}) \label{eotheta}\\
 \theta[^{a+2n}_{b+2m} ](Z|{\bf \Omega}) &= (-)^{am^T} \,\theta[^{a}_{b} ](Z|{\bf \Omega}) \label{shift1}\\
  \theta[^{a}_{b} ]\left(Z+(^n_m)|{\bf \Omega}\right) &= e^{\pi\ii(am^T-nb^T-2nZ^T-n{\bf \Omega}n^T)} \,\theta[^{a+n}_{b+m} ](Z|{\bf \Omega}) \label{shift2}
 \end{align}
 \end{subequations}
 where $(^n_m)$ denotes an half-period as defined in (\ref{halfper}). Eq.~(\ref{eotheta}) is telling us that $\theta[^{a}_{b} ](Z|{\bf \Omega})$ 
 is even/odd in $Z$ if the half-characteristics $[^a_b]$ is even/odd respectively. We will often use the short-hand notation
 \be
   \theta[^{a}_{b} ]\equiv     \theta[^{a}_{b} ]({\bf\Omega})\equiv \theta[^{a}_{b} ](0|{\bf \Omega})
 \ee
 Clearly, by (\ref{eotheta}),    $\theta[^{a}_{b} ]\equiv 0$ if $[^a_b]$ is odd.
 
 Eq.~(\ref{shift1}) implies that, up to a sign, we can reduce  the half-characteristic matrix to take values $0$ or $1$. 
 We denote by $\nu_i$, $i=1,\ldots, 6$, the odd half-characteristics
 \be
\nu_1 = \footnotesize{ \left[
\begin{array}{l}
 0 1    \\
 0 1   \\
\end{array}
\right]}
\quad 
 \nu_2 = \footnotesize{ \left[
\begin{array}{l}
 0 1    \\
 1 1   \\
\end{array}
\right]}
\quad
\nu_3 = \footnotesize{ \left[
\begin{array}{l}
 1 0    \\
 1 1   \\
\end{array}
\right]}
\quad
\nu_4 = \footnotesize{ \left[
\begin{array}{l}
 1 0    \\
 1 0   \\
\end{array}
\right]}
\quad
\nu_5 = \footnotesize{ \left[
\begin{array}{l}
 1 1    \\
 1 0   \\
\end{array}
\right]}
\quad
\nu_6 = \footnotesize{ \left[
\begin{array}{l}
 1 1    \\
 0 1   \\
\end{array}
\right]}
 \ee
Even half-characteristics can be obtained as sums mod 2 of three odd half-characteristics and therefore they can be labeled by  triplets $\{i,j,k\}$. Hence, we can introduce the following shorthand  notation for the theta functions with even half-characteristics evaluated at $Z=0$: 
 \be
 \theta_{ijk}\equiv\theta[\nu_i+\nu_j+\nu_k] 
 \ee
  where the sum is understood mod 2.  More explicitly
\be
\begin{aligned}
& \theta_{123}= \theta[^{1 0}_{ 0 1}]   \qquad 
 \theta_{124}=\theta[^{10}_{00}]\qquad
  \theta_{125}=\theta[^{11}_{00}]\qquad
   \theta_{126}= \theta[^{11}_{ 1 1 }]\qquad
\theta_{134}=\theta[^{ 01}_{ 0 0}]\\
&\theta_{135}=\theta[^{00}_{00}]\qquad
 \theta_{136}=\theta[^{00}_{11}]  \qquad
 \theta_{145} = \theta[^{ 0 0}_{ 0 1}] \qquad
\theta_{146} =\theta[^{00}_{10}]\qquad  
 \theta_{156} = \theta[^{01}_{10}]
 \end{aligned}
\ee
Notice that a triplet of integers and its complementary lead to the same $ \theta_{ijk}$, {\em e.g.}~ $\theta_{123}=\theta_{456}$.

Theta functions evaluated at $Z=0$ transform nicely under modular tranformations:
\be\label{thetamodular}
\theta[{}^{\tilde{a}}_{\tilde{b}}](\tilde{\bf\Omega})=e^{\ii\varphi}\, {\rm det} (C{\bf \Omega}+D)^{1\over 2}\,
\theta[ {}^a_b ]({\bf \Omega})
\ee
where  $\tilde{\bf\Omega}=(A{\bf\Omega}+B)(C{\bf\Omega}+D)^{-1}$ and
\be
\left(
\begin{array}{c}
 \tilde{a}^T    \\
 \tilde{b}^T   \\ 
\end{array}
\right) = 
\left(
\begin{array}{cc}
 D  & -C      \\
 -B &   A  \\ 
\end{array}
\right) \left(
\begin{array}{c}
 a^T    \\
 b^T   \\ 
\end{array}
\right)+
\ft12 {\rm diag} \left(
\begin{array}{c}
 CD^T    \\
 AB^T   \\ 
\end{array}
\right)
\ee
In (\ref{thetamodular}), the phase factor obeys $e^{8\ii\varphi}=1$ and depends on $[^a_b]$ and on the modular transformation.


In terms of theta functions one can write \cite{streng}
\bea
\psi_4 &=& \ft14 \sum_{\delta\in T }   \theta[\delta]^8  \nn\\
\psi_6 &=& \ft14 \sum_{C_3} \epsilon(C_3) \prod_{\delta \in C_3 }  \theta[\delta]^4    \nn\\
\chi_{10} &=& -\ft{1}{2^{14}}  \prod_{\delta\in T}  \theta[\delta]^2 \nn\\
\chi_{12} &=& \ft{1}{2^{17} 3 }\sum_{C_4} \prod_{\delta \notin C_4}   \theta[\delta]^4  \nn\\
\chi_{35} &=& \ft{1}{2^{39} 5^3 \ii }\prod_{\delta\in T}  \theta[\delta]\sum_{C_3'} \epsilon(C_3') \prod_{\delta \in C_3' }  \theta[\delta]^{20}  \label{hs}
\eea
 with $T$ the set of  even characteristics, $C_4$ is the set of quartets of even  characteristics defined as
 \be
 C_4=\{ (\delta_1,\delta_2,\delta_3,\delta_4) \quad {\rm with} \quad  \sum_{i=1}^4 \delta_i=(^{00}_{00})  \} 
  \ee
   There are 15 of such quartets. Finally $C_3$ ($C_3'$) is the set of triplets contained in any element of $C_4$, whose sum is even (odd). There are 60 choices for both cases. The signs $\epsilon(C_3)$ and $\epsilon(C_3')$ in the formulae (\ref{hs}) for $\psi_6$ and $\chi_{35}$ are fixed by modular invariance.
   
It can be useful to introduce the alternative parameters
  \be\label{qqy}
  q_1=e^{2\pi \ii \tau} \qquad q_2=e^{2\pi \ii \sigma}   \qquad y=e^{2\pi \ii \beta} 
  \ee
  and consider the expansions of theta functions evaluated at $Z=0$ for small values of $q_1,q_2$
  \be\label{thetaexp}
 \theta [^a_b ]({\bf\Omega})= 
 q_1^{a_1^2\over 8 }\, 
 q_2^{ a_2^2\over 8} y^{a_1 a_2\over 4   } \,e^{\pi \ii a_i b_i\over 2}
 +q_1^{\left(1-{a_1\over 2}\right)^2}\, 
 q_2^{\left(1-{a_1\over 2}\right)^2} y^{\left(1-{a_1\over 2}\right) \left(1-{a_2\over 2}\right)  }
  \,e^{\pi \ii \left(1-{a_i \over 2}\right) b_i }+\ldots
 \ee
  In particular  for the cusp forms $\chi_{10}$ and $\chi_{12}$ one finds the expansions
 \bea
 \chi_{10} &=& {(1-y)^2\over 4y} \,q_1 q_2+\dots\nn\\
  \chi_{12} &=& {(1+10 y+y^2)\over 12y} \,q_1 q_2+\dots\\
 \eea

\subsubsection{The curve in terms of theta functions}

One can use theta functions to provide a useful parametrization of the hyperelliptic curve in which the 
dependence on the period matrix ${\bf \Omega}$ is manifested. 

First of all, the definition (\ref{sucurve}) of the curve depends on the six branch points $e_i$ while ${\bf \Omega}$ depends on just the three independent parameters $\tau,\sigma,\beta$. One can remove the redundancy of the description 
(\ref{sucurve}) by performing the following transformation 
\be
\hat{x}(P) = \left(\frac{x(P)-e_1}{ x(P)-e_5}\right)\left( \frac{e_3-e_5}{e_3-e_1}\right)
\ee 
 with $x(P)$ the projection of the point $P\in \Sigma$ on the $x$-plane.
Three of the new branch points, denoted by $\xi_i=\hat{x}(P_i)$, have now fixed values $\xi_1=0$, $\xi_3=1$ and $\xi_5=\infty$,
while the other three $\xi_2,\xi_4,\xi_6$ provide non-degenerate information on the curve.

With a proper choice of $a_0$ in (\ref{sucurve}) the curve can be written as
  \be\label{altcurve}
 y^2= \hat{x}(\hat{x}-1)(\hat{x}-\xi_2)(\hat{x}-\xi_4)(\hat{x}-\xi_6)
 \ee
 By using (\ref{abel}), one can construct the functions on the curve $\Sigma$
 \be\label{auxfunct}
f_1(P)=\left( \frac{\theta[^{10}_{10}]( \phi(P)|{\bf \Omega} )}{\theta[^{00}_{01}](\phi(P)| {\bf{\Omega}})}\right)^2\qquad~~~f_2(P)=\left(\frac{\theta[^{01}_{11}]( \phi(P)|{\bf \Omega} )}{ \theta[^{11}_{00}](\phi(P)|{\bf \Omega})}\right)^2
 \ee
 which have both a double zero and a double pole at the branch points  $P_1$  and  $P_5$ respectively, since $\phi(P_i)\sim \sqrt{x-e_i}$.  On the other hand, the map $P\in\Sigma\mapsto \hat{x}(P)\in \mathbb{P}^1$ provided by the double-sheeted description (\ref{altcurve}) have exactly the same zero/pole structure of $f_1(P)$ and $f_2(P)$. This can be seen by using $y\sim  \sqrt{x-e_i}$ as a local coordinate around the branch points.
  This implies that the three functions must coincide up to a multiplicative constant. The multiplicative constant can be fixed by requiring that $f_i(P_3)=1$:
  \be
  \label{xP}
\hat{x}(P)=\left[{\theta[^{10}_{10}]( \phi(P) |{\bf \Omega}) \theta[^{11}_{11}]({\bf \Omega}) \over \theta[^{00}_{01}](\phi(P) |{\bf \Omega}) \theta[^{01}_{00}]({\bf \Omega})}\right]^2= \left[{\theta[^{01}_{11}]( \phi(P)|{\bf \Omega} ) \theta[^{00}_{10}]({\bf \Omega}) \over \theta[^{11}_{00}](\phi(P)|{\bf \Omega}) \theta[^{10}_{01}]({\bf \Omega})}\right]^2
\ee
where the normalization is fixed by using  $\phi(P_3)$ given in (\ref{branchabel}) and the property (\ref{shift2}) of theta functions.
  
By repeatedly  using   (\ref{branchabel}) and  (\ref{xP}), one can compute the remaining $\xi_i$ as functions of the period matrix 
 \be
 \begin{aligned}
 \xi_2 ({\bf \Omega})&={ e_{21} e_{35}\over e_{25} e_{31} } ={ \theta[^{11}_{11}]^2\, \theta[^{10}_{00}]^2 \over \, \theta[^{01}_{00}]^2\, \theta[^{00}_{11}]^2}  ={  \theta_{126}^2 \theta_{124}^2 \over \theta_{134}^2 \theta_{136}^2  }\\
  \xi_4({\bf \Omega}) &= { e_{41} e_{35}\over e_{45} e_{31} } ={ \theta[^{10}_{00}]^2\, \theta[^{00}_{10}]^2\over \theta[^{00}_{11}]^2\, \theta[^{10}_{01}]^2} ={  \theta_{124}^2  \theta_{146}^2\over \theta_{136}^2 \theta_{123}^2  }\\
    \xi_6 ({\bf \Omega})&= { e_{61} e_{35}\over e_{65} e_{31} }  ={ \theta[^{00}_{10}]^2\,\theta[^{11}_{11}]^2\over \theta[^{10}_{01}]^2\, \theta[^{01}_{00}]^2 }={  \theta_{146}^2 \theta_{126}^2 \over \theta_{123}^2 \theta_{134}^2  }   \label{xi234}
    \end{aligned}
 \ee
Alternatively, by rescaling $\hat{x}$ and $y$ in (\ref{altcurve}), we can rewrite it as
\be\label{modcurve}
y^2=x(x-\theta^2_{136}\theta^2_{123}\theta^2_{134})(x-\theta^2_{123}\theta^2_{126}\theta^2_{124})(x-\theta^2_{134}\theta^2_{124}\theta^2_{146})(x-\theta^2_{136}\theta^2_{146}\theta^2_{126})
\ee
This form makes explicit the modular properties of the coefficient of quintic. Indeed shifting $x$ in order to eliminate the $x^4$ term,  (\ref{modcurve}) can be rewriten
as
\be\label{modcurve2}
y^2=x^5+f_6({\bf \Omega})x^3+f_9{({\bf \Omega})}x^2+f_{12}({\bf \Omega})x+f_{15}({\bf \Omega})
\ee
where $f_{k}({\bf\Omega})$ are some modular forms of weight $k$.


  \subsection{Degenerations of the Riemann surface} 
  
 There are two types of degenerations of a genus two curve depending one squeezes a cycle
 homologous to zero or not. Squeezing a cycle non homologous to zero the Riemann surface
 degenerates to a genus one surface with a double point.  In this limit $\tau \to \ii \infty$ or $\sigma\to \ii \infty$. We refer to this degeneration class simply as ``pinching a handle".
 Squeezing a cycle homologous to zero the Riemann surface degenerates into two  genus one surfaces  linked by a long tube and $\beta \to 0$. We refer to this degeneration as ``splitting into 
two genus one surfaces". A complete analytic classification of singular fibers of genus two Riemann surfaces and the definition of their homological monodromies was made by Namikawa and Ueno \cite{Ueno}. 

  The theta functions of a degenerated surface can be always written in terms of elliptic functions. 
For the case in which a handle is pinched by sending $\sigma\rightarrow\ii\infty$ one finds  
\be
\begin{aligned}
 \theta[^{a_1 a_2}_{b_1 b_2}] ({\bf\Omega})&\simeq  \theta[^{a_1}_{b_1} ](\tau) \qquad~~~~~~~~~~~~~~~~~~~~~~~~~~~~~~~~~~~~~~\text{(for $a_2$ even)}\\
  \theta[^{a_1 a_2}_{b_1 b_2}] ({\bf\Omega})&\simeq e^{ \pi i \sigma \over 4}\left(e^{\frac{\pi\ii b_2}{2}}+e^{\pi\ii a_1b_1}e^{-\frac{\pi\ii b_2}{2}}\right)  
  \theta[^{a_1}_{b_1} ]( \ft{\beta }{2}  |\tau)\label{deg0}\qquad~~\text{(for $a_2$ odd)}
  \end{aligned}
\ee 
where
\be
  \theta[^{a}_{b} ]( z  |\tau)= \sum_{n\in \Zint} e^{ \pi \ii \tau \left(n+{a\over 2} \right)^2 
 +2\pi \ii  \left(n+{a\over 2} \right)\left(z+{b\over 2}\right) }  \label{genus1theta}
\ee
are the genus one theta functions. Finally, we recall the following alternative standard notation for theta functions
\be
 \theta_1(z|\tau) \equiv\theta[^{1 }_{1 }](z|\tau),~~ \theta_2(z|\tau) \equiv\theta[^{1 }_{0 }](z|\tau),~~ 
 \theta_3(z|\tau) \equiv\theta[^{0 }_{0 }](z|\tau),~~ \theta_4(z|\tau) \equiv\theta[^{0 }_{1 }](z|\tau)
 \label{thetastand}
\ee
 In the limit $\tau\rightarrow\ii\infty$ one gets a completely analogous formula:
\be
\begin{aligned}
  \theta[^{a_1 a_2}_{b_1 b_2}] ({\bf\Omega})&\simeq  \theta[^{a_2}_{b_2} ](\sigma) \qquad~~~~~~~~~~~~~~~~~~~~~~~~~~~~~~~~~~~~~~\text{(for $a_1$ even)}\\
  \theta[^{a_1 a_2}_{b_1 b_2}] ({\bf\Omega})&\simeq e^{ \pi i \tau \over 4}\left(e^{\frac{\pi\ii b_1}{2}}+e^{\pi\ii a_2b_2}e^{-\frac{\pi\ii b_1}{2}}\right)  
  \theta[^{a_2}_{b_2} ]( \ft{\beta }{2}  |\sigma)\label{deg1}\qquad~~\text{(for $a_1$ odd)}
  \end{aligned}
\ee 

 For the case where the Riemann surface splits into two genus one surfaces one finds
    \be
   \lim_{\beta \to 0} \theta[^{a_1 a_2}_{b_1 b_2}] =  \theta[^{a_1 }_{b_1 }] 
 \theta[^{a_2}_{ b_2}]+{\beta\over 2\pi \ii}  \, \partial_z \theta[^{a_1 }_{b_1 }] (z| \tau)
 \partial_z \theta[^{a_2 }_{b_2 }] (z| \sigma)\Big|_{z=0}+\ldots
 \label{beta0}
 \ee
In the following we describe the details of the two basic degenerations of the genus two curve. We refer the reader  to the Appendix of \cite{Bonelli:2010gk} for a clear exposition of these two kinds of degenerations.

 \subsubsection{Pinching a handle}
 \label{spinching}

  Plugging (\ref{deg0}) into (\ref{xi234}) one finds that the harmonic ratios entering 
  in the hyperelliptic curve at $\sigma\to \ii \infty$
 (or $e_{34}\to 0$)  reduce to
 \be
 \xi_2 =-{\theta_2^2(\tau)\over \theta_4^2(\tau)}{\theta_1^2(\ft{\beta}{2}|\tau)\over \theta_3^2(\ft{\beta}{2}|\tau)} \qquad    \xi_4=1 \qquad \xi_6=-{\theta_4^2(\tau)\over \theta_2^2(\tau)}{\theta_1^2(\ft{\beta}{2}|\tau)\over \theta_3^2(\ft{\beta}{2}|\tau)} \label{oneh}
 \ee
where we used the standard notation (\ref{thetastand}) for the genus one theta functions. 
 The curve reduces to the form
   \be
   \tilde y^2=x(x-\xi_2)(x- \xi_6)
      \qquad ~~~~~~~~~~~\tilde y={y\over (x-1)}
      \ee
which corresponds to a genus one curve with a double point at $x=1$ and harmonic ratio
\be
 \frac{\xi_2}{\xi_6}={\theta_2^4(\tau)\over \theta_4^4(\tau)}
\ee
 Other handle degenerations are related to this by the action of the modular group.

\subsubsection{Pinching two handles}
\label{twopinch}
  The limit where both handles are pinched can be found from (\ref{oneh}) sending $\tau\to \ii \infty$. This leads to $\xi_2\to 0$ and degenerates the curve to an irreducible rational curve with two ordinary double points (in $x=0,1$)
 \be
   \tilde y^2=(x- \xi_6)
      \qquad ~~~~~~~~~~~\tilde y={y\over x(x-1)}
      \ee
  with
  \be
  \xi_6= - \sin^2 {\pi \beta\over 2}
  \ee

%
   
 \subsubsection{Spitting into two genus one surfaces}

\label{splitting}

 In the limit $\beta\to 0$, using (\ref{xi234}) and  (\ref{beta0}) one finds  
the harmonic ratios  
   \be
  \label{degratios}
  \xi_2=\frac{\beta^2}{a_2}~~~~~~~~~~~~\xi_4(\sigma)= {\theta_3^4\over \theta_4^4}(\sigma) ~~~~~~~~~~~~\xi_6=\frac{\beta^2}{a_6}
  \ee   
  with
\bea
  a_2 &=& -\frac{ 4 }{\pi^2\,\theta_3^4(\sigma)\,  \theta_2^4(\tau)}~~~~~~~~~  a_6  =-\frac{ 4}{\pi^2\,
 \theta_3^4(\sigma)  \theta_4^4(\tau)} 
\eea
obtained by using the relation $\partial_{z}\theta_1(z|\tau)|_{z=0}=\pi \,\theta_2 \theta_3\theta_4 (\tau)$.
  Away from $x= 0$  the Riemann surface is then  described by the elliptic curve
     \bea
    y_1^2 &=&   x (x-1)\left(x- {\theta_3^4\over \theta_4^4}(\sigma)  \right)      \qquad ~~~~~~~~~~~~~~~~~~~~ y=x\, y_1
    \label{y1}
    \eea
 On the other hand,  near $x= 0$, one can write
    \be
    x={\beta^2\over \tilde x  }  \qquad~~~~~~~~~~~~~~~ 
 y=  \frac{y_2\,\beta^3}{ \tilde x^2}\sqrt{ \xi_4\over\,a_2\,a_6}  
\ee
to bring  the curve into the elliptic form
 \be
y_2^2 =  \tilde x ( \tilde x-a_2)( \tilde x -a_6)   
 \ee
 with harmonic ratio
 \be
 \frac{a_6}{a_2}={\theta_2^4(\tau)\over \theta_4^4(\tau)}
 \ee
 Summarizing, at $\beta\to 0$, the Riemann surface splits into two genus one curves 
 (near and far away from $x= 0$) with complex structure parameters given in terms 
 of $\sigma$ and $\tau$ respectively.
 We notice that the limit $\beta\to 0$  corresponds to sending $e_1,e_2,e_6$
  together as follows from $\xi_2=\xi_6=0$.

 \subsection{An example of hyperelliptic fibration}   
  \label{AppPerios}
 In this section we determine the degenerations and holonomies of the fiber period
 matrix for a simple choice of hyperelliptic curve. We take 
 \be
 y^2=(x^3-z)^2-g^2 (x^3-m^3)^2=\prod_{i=1}^6 (x-e_i(z)  ) \label{su30}
 \ee
with $g,m$ some constants. The six branch points are given by
\be
e_{2k}(z)=w^{k-1} \left(z+ g\, m^3\over 1+ g \right)^{1\over 3}  \qquad e_{2k-1}(z)=w^{k-1} \left( z- g\, m^3\over 1- g \right)^{1\over 3}
 \ee
with  $ w=e^{2\pi i \over 3}$, $k=1,..3$.
  The curve (\ref{su30}) defines a genus two surface at each point $z$. The fiber degenerates
   at the points
  \bea
  z=m^3  && \qquad   e_{2k}=e_{2k-1}=w^{k-1} \, m      \nn\\
  z=g\, m^3  && \qquad   e_{2k-1}=0  \nn\\
  z=-g\, m^3  && \qquad   e_{2k}=0  
  \eea  
where some of the branch points collide.   The monodromies around these points can be derived from the period matrix in the nearby   of the degeneration point. Let us consider for example the curve near $z=m^3$. 
   The period matrix ${\bf \Omega}(z)$ of the Riemann surface is given by (\ref{MN}). 
   
   At  $z= m^3$ the $\gamma$-cycles shrink to zero and the corresponding integrals 
boil down  to a residue 
\be
M^a{}_{b}   ={1\over 2\pi i  \sqrt{1-g^2}} \oint_{\gamma^b} { x^{a-1} dx \over  (x^3-m^3) }=
-{w^{a (b-1)} m^{a-3}\over  3\sqrt{1-g^2} }
\ee
On the other hand  for the $\tilde{\gamma}$-integrals one finds
\bea
N^a{}_{1} &=& {1\over  \pi i \sqrt{1-g^2}} \left(\int_{e_2}^{e_3} +\int_{e_4}^{e_5}\right) { x^{a-1} dx \over  (x^3-z) }\simeq { m^{a-3} (-1+w^{-a} )\over   3\pi i\sqrt{1-g^2}} \, \log\left(1-{z\over m^3}     \right) \nn\\
N^a{}_{2} &=& {1\over  \pi i \sqrt{1-g^2}} \int_{e_4}^{e_5} { x^{a-1} dx \over  (x^3-z) }\simeq { m^{a-3} (-)^a \ii  \over  \pi i \sqrt{3(1-g^2)}} \, \log\left(1-{z \over m^3}     \right) 
\eea
 resulting into
 \be
{\bf  \Omega}(z) \simeq {1\over2\pi \ii} \log\left(1-{z\over m^3}     \right)  
\left(
\begin{array}{cc}
  4&2     \\
  2  &  4   \\   
\end{array}
\right)
 \ee
near $z= m^3$. Going around $z=m^3$ one finds the monodromies
\bea
\tau \to \tau+4 \qquad    \sigma\to \sigma+4 \qquad \beta\to \beta+2
\eea
  A similar analysis can be done for the holonomies at the degeneration points $z=\pm g m^3$. 
  In particular for $g\to 0$ when the two points come together one finds the monodromies
  associated with O-planes at $z=0$ with charges compensating for D-branes at $z=m^3$.

\end{appendix}


\newpage

\providecommand{\href}[2]{#2}\begingroup\raggedright\endgroup


\end{document}